\documentclass[aps,prd,twocolumn,amsmath]{revtex4}
 \usepackage{natbib}
\usepackage{graphicx}% Include figure files
\usepackage{dcolumn}% Align table columns on decimal point
\usepackage{bm}% bold math
\usepackage{amsmath}
\usepackage{multirow}
\usepackage{float}
\usepackage{amssymb}

\begin{document} 

\title{Moments of multiplicity distributions using Tsallis statistics in leptonic and hadronic collisions }
\author{S. Sharma}
\author{M. Kaur}

\affiliation{Department
of Physics, Panjab University, Chandigarh -160 014, India}

\date{\today}

\email{sandeep.sharma.hep@gmail.com} 

\date{\today}% It is always \today, today,
\begin{abstract}
Moments play a crucial role in investigating the characteristics of charged particle multiplicities in high energy interactions. The success of any model which describes the multiplicity data can be understood well by studying the normalised and factorial moments of that distribution.~The Tsallis model is one of the most successful models which describes the multiplicity spectra, transverse momentum ($p_T$) spectra very precisely in high energy interactions.~In our previous work we have used the Tsallis $q$-statistics to describe the multiplicity distributions in leptonic and hadronic collisions at various energies ranging from 14 GeV to 7 TeV.~In the present study we have extended our analysis for calculating the moments using the Tsallis model for $e^+e^-$ interactions at $\sqrt{s}$ = 91 to 206 GeV from the LEP data and for $pp$ interactions at $\sqrt{s}$ = 0.9 to 7 TeV in various pseudo-rapidity intervals from the CMS data at LHC. By using the Tsallis model we have also calculated the average charged multiplicity and its dependence on energy.~It is found that the moments and the mean multiplicities predicted by the Tsallis model are in good agreement with the experimental values.~We have also predicted the mean multiplicity at $\sqrt{s}$ = 500 GeV for $e^+e^-$ collisions and at $\sqrt{s}$ = 14 TeV for $pp$ collisions in extreme pseudo-rapidity interval, $|\eta|$ $<$ 2.4
\end{abstract}  

\maketitle

\section{Introduction}
In high energy collisions, particles are made to collide with relativistic momenta much greater than their rest masses resulting in the production of large number of particles in final state \cite{m1} from a variety of processes.~These collisions can be  hadronic, leptonic  or heavy-ion interactions; summarized in the form of reaction, for leptonic collision as $l$-$l \rightarrow X$ \cite{lc}, where $l$ is the lepton or for hadronic collisions as $h$-$h \rightarrow X$  \cite{lchc}, where $h$ is the hadron or for hadron-nucleus collision as $h$-$A \rightarrow X$ \cite{ha}, with $A$ being the nucleus.~$X$ in the final state of these reactions represents any number of particles, produced due to the gluon-gluon, quark-quark and quark-gluon interactions between the constituent quarks and gluons of the colliding particles.~The produced particles can be the baryons (qqq state), mesons ($\bar{q}q$ state) or leptons.~Simplest but the most significant observation to describe the mechanism of particle production is the observation of charged particle multiplicity \cite{m5, m6} and the distribution of number of particles produced, known as multiplicity distribution \cite{imdremin}.~Multiplicity distribution, MD, also carries important information about the correlations of particles produced, thus providing a very fine way to inquest the dynamics of the quark-quark, gluon-gluon and  quark-gluon interactions.

Collision or interaction of two particles is generally described in terms of cross-section which is calculated by measuring the number of particles produced.~The cross section essentially gives the measure of the probability of production of particular number of particles.~The multiplicity distribution is defined in terms of probability by the formula; 
%\bigskip
\vspace{-0.5cm}
\begin{center}
	
	\begin{equation}
 P_N = \frac{\sigma_N}{ \sigma_{total}} = \frac{N_{ch}}{N_{total}}
	\end{equation}

\end{center}
%\vspace{-0.3cm}
%\bigskip

where $\sigma_N$ is the cross section for production of `$N$' number of particles  and $\sigma_{total}$ represents the total cross section of interaction at center of mass energy $\sqrt{s}$.~Experimentally this probability, $P_{N}$, is obtained from the number of charged particles produced at specific multiplicity, $N_{ch}$ and  the total number of particles  produced during the collisions, $N_{total}$.~The multiplicity distribution, MD, obeys conventional Poisson distribution \cite{p1} if there is no correlation between the particles produced i.e. particles produced are exclusive and independent of each other \cite{fs0}.~The presence of any kind of correlation amongst the particles leads to the deviation from Poissonian form.~Higher order moments and its cumulants are the precise tools to study the correlation between the particles produced in these interactions \cite{cor1, cor2}.

In the last few years Tsallis model \cite{ts} has been used successfully in describing the MDs in hadronic and leptonic collisions for different collision energies.~Recently we have analysed the $e^{+}e^{-}$, $pp$ and $\bar{p}p$ collisions at different energies by using the Tsallis model \cite{shm1, prd2, prd1, shm}.~In the present study we use the Tsallis approach to measure correlations between the particles produced in both leptonic and hadronic interactions at energies ranging from few GeV upto the LHC energies.~Additionally the dependence of the average multiplicity on centre of mass energy is also studied.~Results from the Tsallis model are compared with experimental values.~In section II we give brief description of moments and the formulation to calculate the higher order moments.~Section III gives the details of the data used and results obtained from the Tsallis model and its comparison with the experimental values.~Discussion and conclusion are presented in section IV.

\section{Moments}
 Multiplicity distributions at low energies $\sim$ 10 GeV for leptonic and hadronic collisions such as $e^{+}e^{-}$ or $pp$, could be described very well using Poisson distribution \cite{p1, p2}.~In such cases, dispersion, defined by $D = \sqrt{<N^{2}> - <N>^{2}}$, is related with the average multiplicity $<N>$.~The multiplicity distributions exhibited a broader width at higher energies showing the significant deviation from the Poissonian form.~The correlation in the particles produced during the collisions was found to be  responsible for the deviations.~The shape of multiplicity distribution can be described well using the assumption that energy dependence of multiplicity distribution at higher energies could be formulated using the average multiplicity.~To explain the energy dependence of multiplicity, Koba, Nielsen and Olesen \cite{KNO} in 1972 proposed the scaling relation for multiplicity distributions known as KNO scaling.~It is the theory of universal scaling  for multiplicity distributions in the asymptotic limit of energy.~The energy dependence of the dispersion defined by relation $D$ $\propto$  $<N> $ implied the compliance of KNO scaling.~But few years later violation of KNO scaling was observed by UA5 collaboration while analysing the multiplicity data at $\sqrt{s} = 540$ GeV \cite{use540} obtained from $\bar{p}p$ collisions.~Later on it was shown by the collaboration that KNO scaling was violated even at  $\sqrt{s} = 200$ GeV \cite{use200}.~Higher order moments and its cumulants are the precise tools to study the correlations between the particles produced in collisions \cite{pp, c2}.~The departure from independent and uncorrelated production of particles can be measured well using the factorial moments, $F_m$ \cite{c3}.~Not only the correlation between the particles but the violation or holding of KNO scaling at higher energies can also be studied and understood correctly by using the normalized moments of order $m$, $C_m$ \cite{wb1}.~These moments are defined as; 

\vspace{-0.5cm}
\begin{center}
 \begin{equation}
C_m = \frac{<N^m>}{<N>^m}  
   \end{equation}
 \end{center}
 \vspace{-0.5cm}
 
\begin{center}
 \begin{equation}
F_m = \frac{<(N(N-1)....(N-m+1))>}{<N>^m}
\end{equation}
 \end{center}
 \vspace{-0.2cm}
 
 The factorial moments and their cumulants, $K_m$, are near to precise in defining the tail part of distribution where events with multitude of particles give a meaningful contribution. The factorial moments and cumulants are related to each other by the relation;
 \vspace{-0.7cm}
\begin{center}
 \begin{equation}
F_m = \sum _{i=0} ^{m-1} C _{m-1} ^i K_{m-i} F_i
\end{equation}
 \end{center}
 %\vspace{-0.5cm}
  
Factorial moments exhibit the features of any kind of correlation present between the particles and cumulants of order $m$ illustrate absolute $m$-particle correlation which can not be brought down to the lower order correlation.~In other words, if all $m$ particles are related to each other in $m^{th}$ order of cumulants, then it can not be divided in to disconnected groups i.e. $m$ particle cluster can not be split in to smaller clusters.~These moments and their dependence on energy $\sqrt{s}$ help in improving, redefining and rejecting various Monte-Carlo or statistical models which can be used in describing the production of particles at high energies. 

\subsection{Tsallis Distribution}
Tsallis statistics \cite{ts} uses the concept of non extensive nature of entropy which is the modification of the usual Boltzman-Gibbs \cite{bz} and is given by;
%\vspace{-0.5cm}
\begin{equation}
S = \frac{1-\sum_{a}P_{a}^q}{q-1}\, 
\end{equation}
%\vspace{-0.5cm}
where  $P_a$  is the probability associated with microstate $a$ and sum of the probabilities over all microstates is normalized;
   $\sum_{a}P_{a}=1$.~For the entropic index $q$ with value $q>1$, $1-q$ measures the departure of entropy from its extensive behaviour.

The probability distribution function in the case of Tsallis $q$-statistics is defined using the partition function Z as,
 \vspace{-0.2cm}
\begin{equation}
P_N = \frac{Z^{N}_q}{Z}
\end{equation}
where Z represents  the total partition function and $Z^{N}_q$ represents partition function at a particular multiplicity. For N particles, partition function can be written as;
 \vspace{0cm}
\begin{equation}
Z(\beta,\mu,V) = \sum(\frac{1}{N!})(nV-nv_{0}N)^{N}
\end{equation}
$n$ represents the gas density, V is the volume of the system and $v_{0}$ is the excluded volume.~The generating function of the distribution plays an important role in providing the physical information of the multiplicity distribution.~The generating function for multiplicity distribution is related to the probability as; 
%\: \: = \: \: \frac{1}{Z_q} \: \sum_n Z_q^{(n)} t^n
\vspace{-0.5cm}
\begin{center}	
\begin{equation}
G(t) \: = \: \sum_{N=0}^{\infty} \: P_N t^N 
\end{equation}
 \end{center}
and can be obtained by using the expression of probability distribution function as given by;
 \vspace{-0.5cm}
  \begin{center}
  	
%\begin{equation}
\begin{multline}
G(t) \approx \exp(t-1) V n [1 + (q-1) \lambda (V n \lambda - 1) -2 v_0 n ]\\
+ (t-1)^2 (V n)^2 [(q-1) \frac{\lambda ^2}{2} - \frac{v_0}{V} ]
\end{multline}
 \end{center}
 %\vspace{-0.5cm} 

The Tsallis probability generating function has the same form as that of Negative Binomial distribution ($G_{NBD} = [1 - \frac{<N>}{k} (t-1) ]^{-k}$ = $\exp [<N> ( t-1)]$) with average of number of particles $\bar{N}$ for Tsallis probability as;
%  \vspace{-0.1cm}
 \begin{equation}
 \bar{N}=Vn[1 + (q-1)\lambda(V n\lambda-1)- 2v_0n], 
 \end{equation} 
 where $\lambda$ is related to the temprerature through parameter $\lambda$ as; 
  \begin{center}
  \vspace{-0.4cm}
 \begin{equation}
\lambda (\beta, \mu)  = - \frac{\beta}{n} \frac{\partial n}{\partial \beta}
\end{equation}
 \end{center}
  
  More details about the calculations can be found in references \cite{ag, shm}

\subsection{Moments of the Tsallis Distribution }
 
The normalized moments of order $m$ of the Tsallis disribution can be calculated through the average number of particles using $C_m = \frac{<N^m>}{<N>^m}$, where average multiplicity $\bar{N}$ = $<N>$.~The factorial moments are defined as;
 \vspace{-0.3cm}
\begin{center}
 \begin{equation}
F_m = \frac{<(N(N-1)....(N-m+1))>}{<N>^m}
\end{equation}
 \end{center}
 \vspace{-0.4cm}
 
 \begin{center}
 \begin{equation}
F_m = \Big(\frac{1}{<N>^m}\Big) \frac{d^{m} G(t)}{dt^m}
\end{equation}
 \end{center}

where $G(t)$ is the generating function of the Tsallis distribution defined by equation (9).~The factorial moments are related to the normalised moments and can be written in the terms of $C_m$. The first five factorial moments are;
  %
  %\begin{center}
 %\begin{equation}
\begin{multline}
  F_2 = C_2 - \frac{C_1}{<N>} \hspace{7cm} \\ 
  F_3 = C_3 - 3\frac{C_2}{<N>} + 2\frac{C_1}{<N>^2} \: \: \: \hspace{3.2cm}\\
   F_4 = C_4 - 6 \frac{C_3}{<N>} + 11\frac{C_2}{<N>^2} - 6\frac{C_1}{<N>^3} \hspace{1.5cm} \\
  F_5 = C_5 - 10 \frac{C_4}{<N> } + 35\frac{C_3}{<N>^2} -  50\frac{C_2}{<N>^3} +  24\frac{C_1}{<N>^4} \\
  \end{multline}
 %\end{equation}
% \end{center}
\vspace{-0.2cm}
\section{ Results}

Experimental data of proton proton collisions from CMS experiment at Large Hadron Collider and the data of $e^+ e^-$ annihilation at different collision energies from the OPAL and L3 experiments are analyzed.~The $pp$ data are analysed at $\sqrt{s}$ = 0.9,~2.34,~7~TeV in the restricted pseudo-rapidity windows of $|\eta|< 0.5,1.0, 1.5, 2.0, 2.4$ \cite{cms}.~The leptonic data from the L3 and OPAL experiments  at $\sqrt{s}$ = 91 to 206 GeV \cite{opal91, opal133, opal161, opalt, l3} in the full phase space are analysed.~Various analyses on the multiplicity distrubtions using these data have been done by us previously and results can be found in the references  \cite{shm1, prd2, prd1, shm}.~In the following sections results of the moments and average multiplicities  obtained using the Tsallis statistics at different energies are discussed%\vspace{-1cm} 

\subsection{Average Multiplicities}

The energy dependence of mean charged multiplicity $<N>$ is expected to reflect the underlying particle production process. ~A number of phenomenological models have been proposed to describe the behaviour of mean charged multiplicity with energy.~One of the  most widely accepted relations which describes the multiplicity as a function of energy $\sqrt{s}$ is \cite{er0};

\vspace{-0.8cm}
\begin{center}
\begin{equation}
<N> = a + b \: ln(\sqrt{s}) + c \: ln^{2}(\sqrt{s})
\end{equation}
\end{center}

We calculate the values of average charged multiplicity from the Tsallis model and compare with the experimental values for $e^+e^-$ collision data and for $pp$ collision data. The $<N>$ calculated from the Tsallis model are given in Tables I $\&$ II.~The values are found to be in good agreement with the experimental values taking the errors in account.

Figure 1 shows the comparison of $<N>$ values from the data and the Tsallis model for $e^+$ $e^-$ collisions at different centre of mass energies, with $\sqrt{s}$ expressed in GeV.~The data are also fitted to the equation (15) as given below; 

\vspace{0.3cm}
For data: 
\vspace{-0.2cm}
\begin{equation}
 <N> \: = \: 176.74   - 70.52 (ln \sqrt{s}) + 7.99  (ln\sqrt{s})^{2}
\end{equation}

 For Tsallis Model:
  \vspace{-0.2cm}
 \begin{equation}
  < N > \: = \: 134.85 - 53.11 (ln \sqrt{s}) + 6.183 (ln\sqrt{s})^{2}
\end{equation}

In case of $pp$ collision data the extreme pseudo-rapidity regions, $|\eta|$ $<$ 0.5 and 2.4 are chosen because of availability of  $<N>$ values for these  pseudo-rapidities only.~Figure 2 shows the comparison of $<N>$ values from the data and the model at $|\eta|$ $<$ 2.4.~In this case the empirical relation describing the dependence of $<N>$ on the centre of mass energy, with $\sqrt{s}$ in TeV takes the form as;
 
%\vspace{-1.5cm}
For  data:
\vspace{-0.2cm}
\begin{equation}
 <N> \:  = \: 18.77 + 4.39 (ln \sqrt{s}) + 0.845 (ln \sqrt{s})^2
\end{equation}	  
	%\end{center}
%\vspace{-1.5cm}

For  Tsallis  Model: 
\vspace{-0.2cm}
	%\begin{center}
	\begin{equation}
  <N>\: =  \:19.35  + 3.874 (ln \sqrt{s}) + 1.146 (ln \sqrt{s})^2
	\end{equation}
	%\end{center}
 Using Tsallis model the average multiplicity is predicted for $e^{+} e^{-}$ interactions at $\sqrt{s}$ = 500 GeV in full phase space and for $pp$ interactions at 14 TeV for pseudo-rapidity range, $|\eta|$ $<$ 2.4.    
 
For $e^{+} e^{-}$ interactions at  $\sqrt{s}$ = 500 GeV the value of average multiplicity $<N>$ is found to be 43.53 $\pm$ 3.79. Where as for  $pp$ collisions the value of $<N>$ at  $\sqrt{s}$ = 14 TeV at $|\eta|$ $<$ 2.4 is found to be 36.18 $\pm$ 3.21. 
 \subsection{Moment Analysis}
The Tsallis gas model has been used to calculate the moments in order to understand the correlation of the final particles produced during the interaction process.~The Tsallis distribution calculated from equation~(6) is fitted to the experimental data on multiplicity distributions at each of the energies.~The multiplicity distribution obtained from the Tsallis model is then used to calculate the moments of the distribution using equations ~(2)~and~(3).~Figure~3 shows the dependence of normalised and factorial moments on the centre of mass energy $\sqrt{s}$, calculated by using i) the Tsallis model and also ii) experimental distributions for $e^+$ $e^-$ data.~The values of these moments are compared with the experimental values and are listed in Tables III and IV.~It is observed that the $F_m$ and $C_m$ moments in each case is non-zero and remains nearly constant with energy.

Moments are also calculated for the $pp$ collisions using the CMS data at different pseudo-rapidity intervals.~The dependence of normalised moments, $C_m$ and factorial moments, $F_m$ on the  pseudo-rapidity, $|\eta|$ at a given energy and dependence of $C_m$ and $F_m$  on energy, $\sqrt{s}$ at a given pseudo-rapidity interval are shown in figures 4-7.~Figures 4-5 show the dependence of the normalised and factorial moments on the pseudo-rapidity intervals at the given energy \cite{cms}.~The value of $C_m$  decreases with increase in the pseudo-rapidity interval at a given energy.~This decrease is clearly visible for $C_5$ because of its large values.~But in the case of factorial moments, value of $F_m$ remains the same within limit with increase in pseudo-rapidity interval as shown in figure~5.~Moments obtained from the Tsallis distributions at these pseudo-rapidity intervals at various  energies are compared with the CMS experimental values and are given in Table~V.~It is found that at each set of pseudo-rapidity intervals the values of both the moments $C_m$ and $F_m$ increase with increase in centre of mass energy, $\sqrt{s}$ as shown in figures~6-7.~The values of the moments obtained using the Tsallis distribution are found to be in good agreement with the experimental values in both the cases of leptonic and hadronic collisions.~In both kinds of interactions it is found that the factorial moments are larger than unity which indicates the presence of correlations amongst particles and deviation from the independent production mechanism. 
	
In the case of leptonic interactions it is found that the values of normalised moments $C_m$ and factorial moments $F_m$  are independent to the centre of mass energy and remain constant with increase in $\sqrt{s}$ within energy range of 91 to 206~GeV.~However in the case of hadronic interactions both type of moments increase with increase in $\sqrt{s}$ whereby the range of $\sqrt{s}$ extends from 900~GeV to 7~TeV.~These results clearly point toward an understanding of behaviour of produced particles.~This also indicate the violation of KNO scaling at larger energies.~But no violation of KNO scaling is observed at lower energies as indicated by the study of leptonic interactions.  

\begin{figure}[H]

\includegraphics[width=3.8 in, height= 2.5 in]{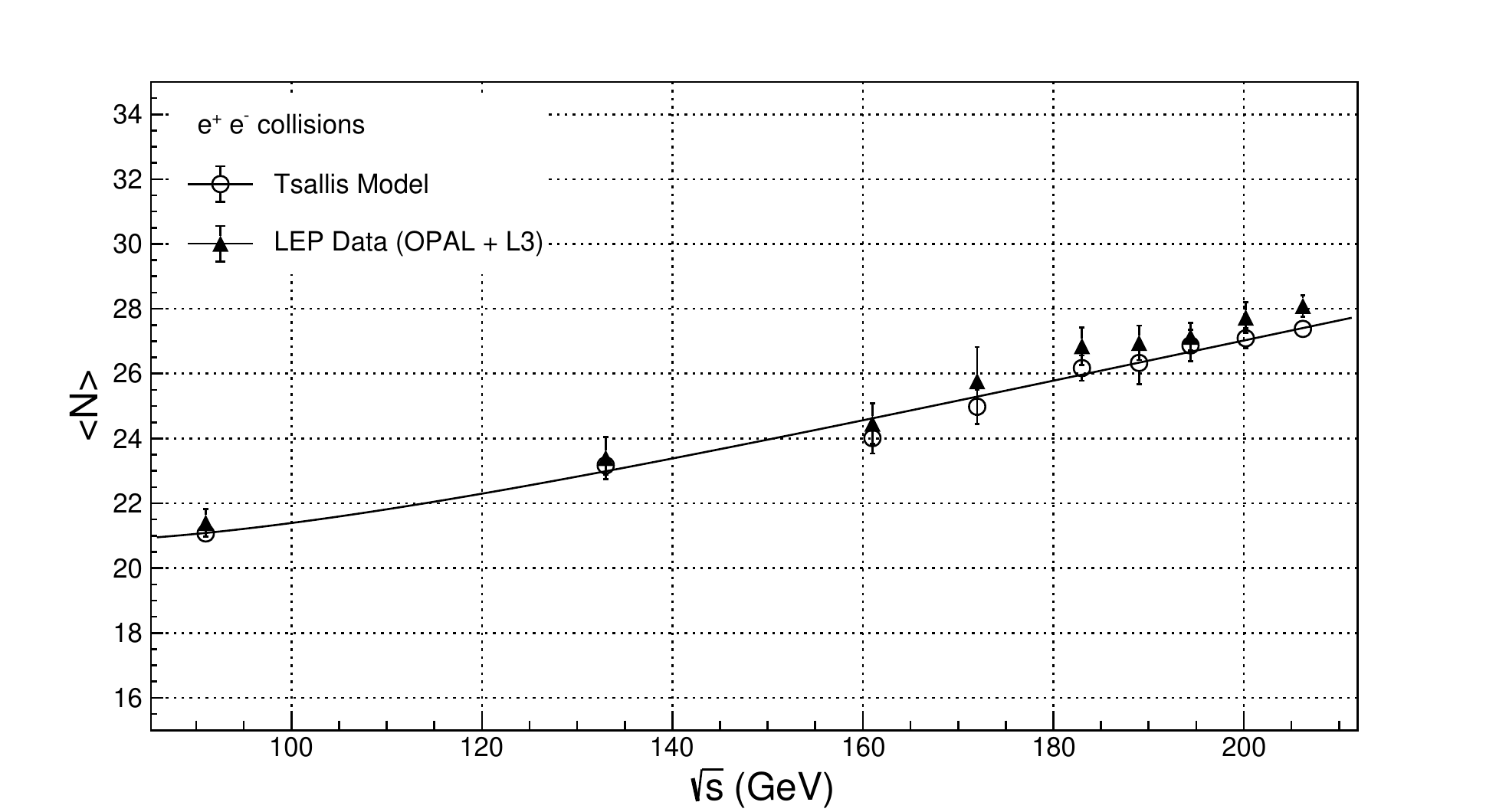}
\caption{Dependence of the average multiplicity $<N> $ on the centre of mass energy, $\sqrt{s}$ for $e^+e^-$ collisions and comparison with experimental values. The solid line corresponds to the equation (16)}
\end{figure}

\vspace{-0.2cm}
\begin{figure}[H]

\includegraphics[width=3.8in , height= 2.5 in]{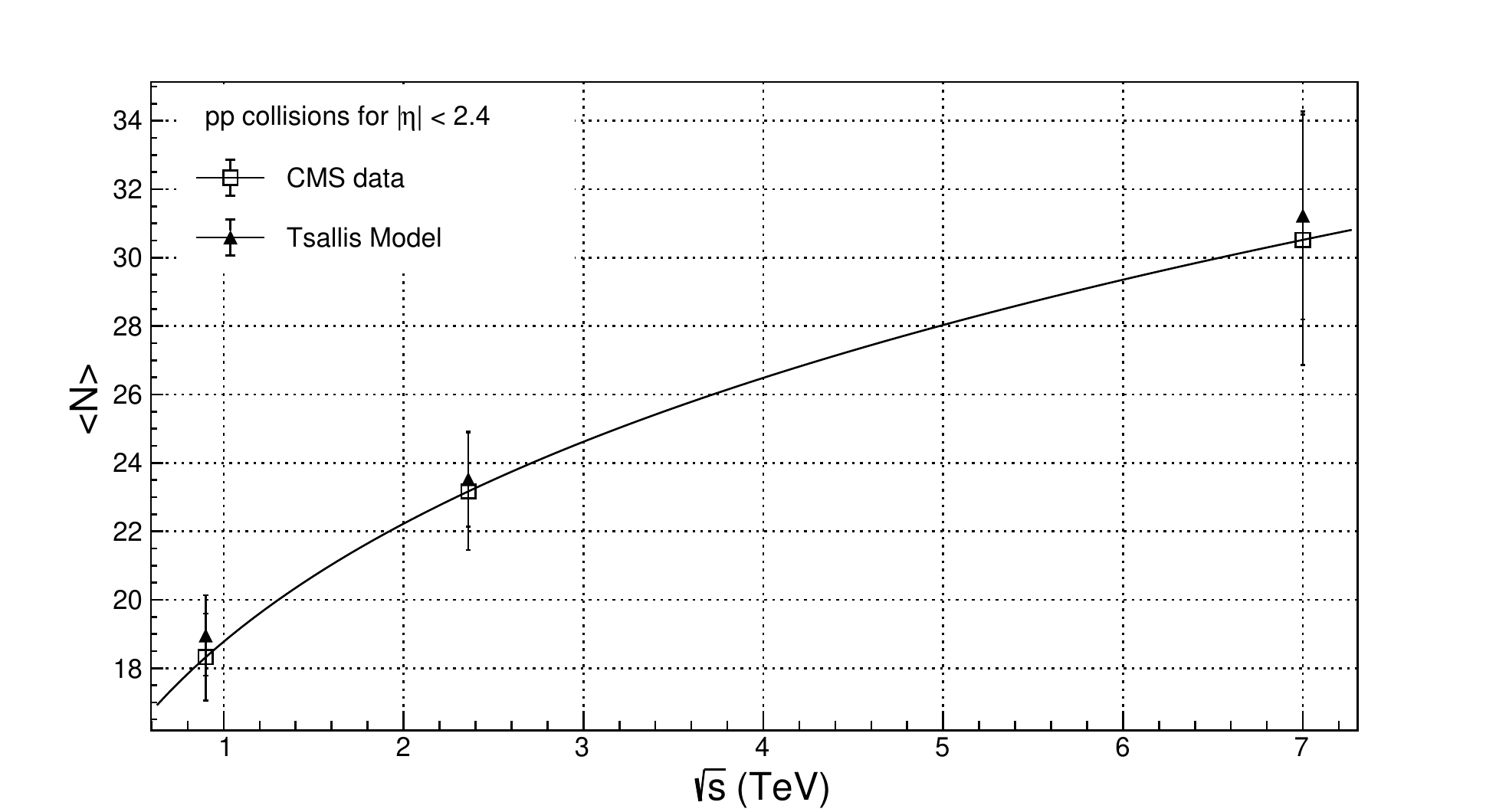}
\caption{Dependence of the average multiplicity on the centre of mass energy. The values from the Tsallis model is compared with the CMS experimental values. The solid line is the fit for the Tsallis model from equation~(18).}
\end{figure} 
 \section{CONCLUSION}
Detailed analysis of the data on electron-positron annihilation at energies $\sqrt{s}$ = 91 to 206~GeV and proton proton collisions at  $\sqrt{s}$= 0.9 to 7~TeV in various pseudo-rapidity regions has been done by using the Tsallis distribution.~The particle production in such interactions is not un-correlated.~The dynamical fluctuations arising due to random cascading processes in particle production can lead to correlations amongst the particles.~The study of higher-order moments of the distributions serves as a very important tool to understand these correlations.~The deviation from independent production can be understood if the factorial moments are larger or smaller than unity.  ~The violation or holding of KNO scaling at higher energies can also be studied and understood correctly by using the normalized moments.~The  KNO scaling implies the energy independence of these moments whereas energy dependence of these moments reflects the KNO scaling violation.~The normalised and factorial moments have been calculated using the Tsallis model and compared with the experimental values.~The obtained values of moments are found to be in good agreement with the experimental values, within experimental uncertainties.~The  values obtained from the Tsallis gas model confirm the violation of KNO scaling as observed in the experimental values at higher energies but no such violations are observed at the lower energies.~Also the average multiplicity values calculated from the Tsallis model have been compared with the experimental values and found to be in good agreement with them.~Using the Tsallis model we have predicted the values of average multiplicties for $e^+e^-$ collisions at $\sqrt{s}$ = 500~GeV and for $pp$ collisions at $\sqrt{s}$ = 14~TeV in pseudo-rapidity region $|\eta|<$ 2.4.~In one of the previous studies the value of average multiplicity at $\sqrt{s}$ = 13~TeV at pseudo-rapidity region $|\eta|$ $<$ 2.4 was predicted by using the Weibull model by A. Pandey et al. \cite{wb1}.~The $<N>$ value predicted by the Tsallis model at $\sqrt{s}$ = 14~TeV at pseudo-rapidity interval $|\eta|$ $<$ 2.4 is found to be consistent with the value predicted by the Weibull model \cite{wb1}.~The study of moments of multiplicity distributions and dependence of average multiplicity on the energy provides interesting features of particle production and helps in understanding the mechanism of particle production at higher energies.~It will be interesting to study the behaviour of particles produced at higher LHC energies ($\sqrt{s}$ $>$ 13~TeV) in future with the Tsallis model.

\begin{figure}[H]
\includegraphics[width=3.5 in , height= 2.3 in]{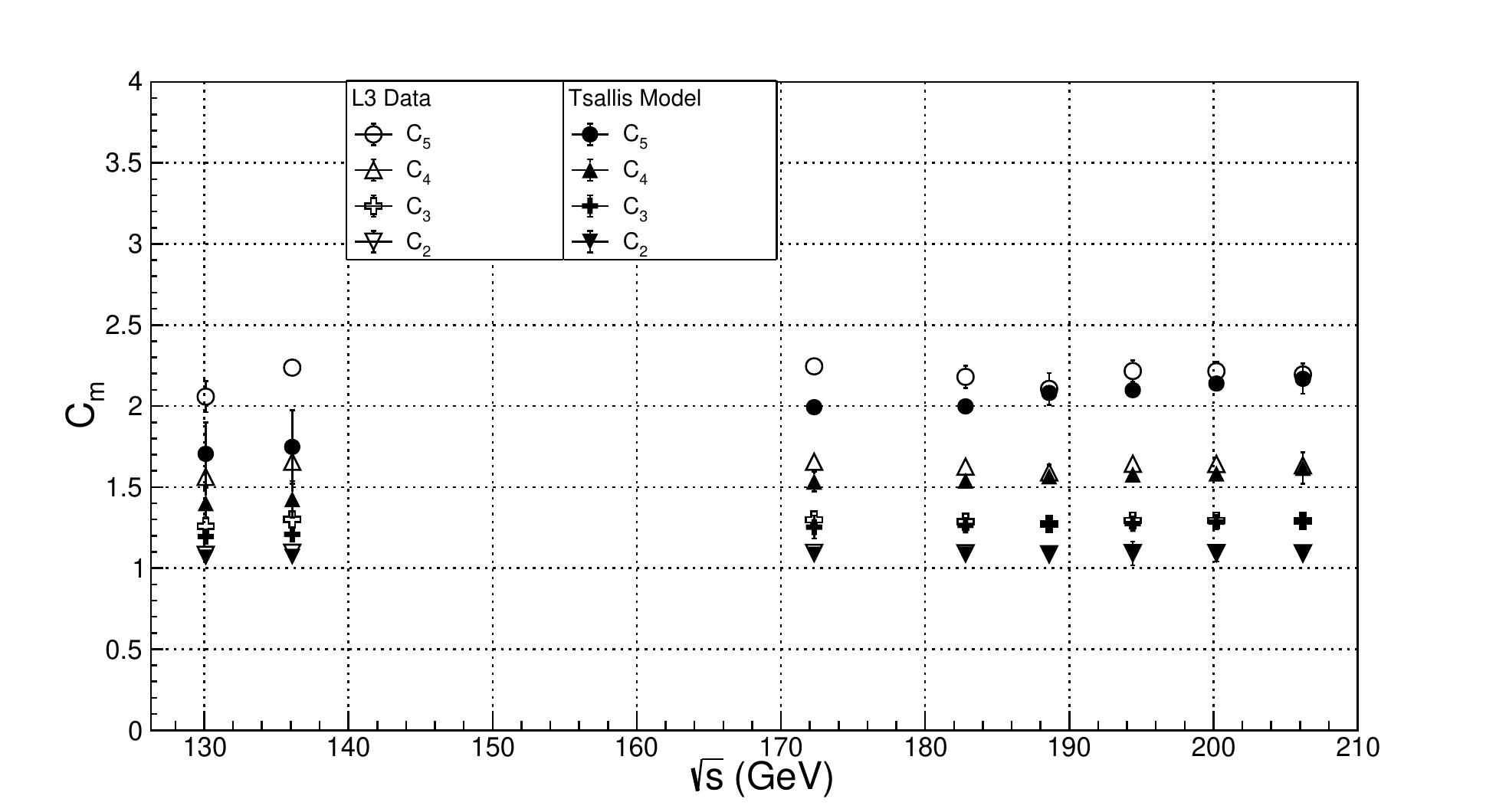}
\includegraphics[width=3.5 in , height= 2.3 in]{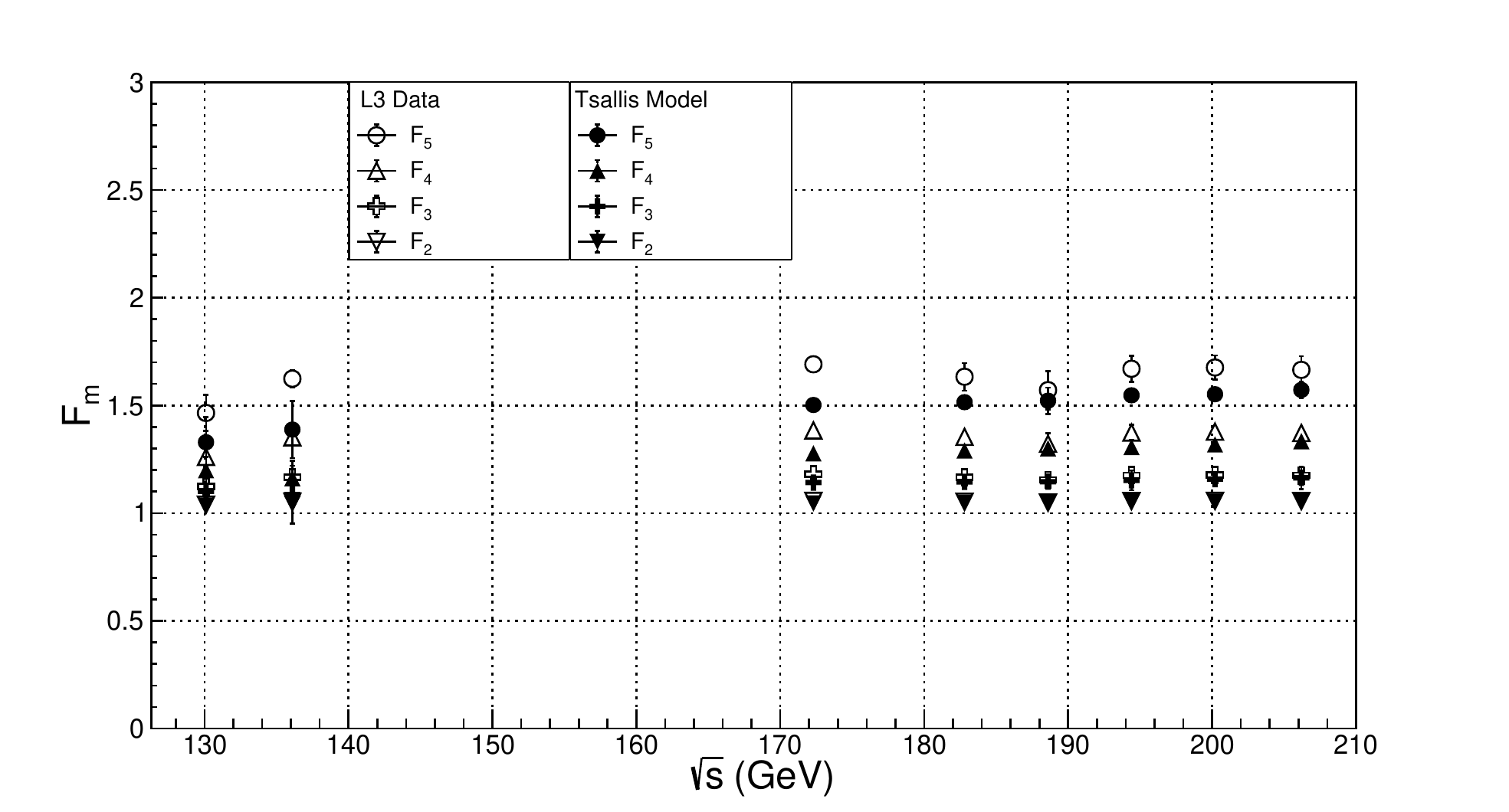}
\includegraphics[width=3.5 in , height= 2.3 in]{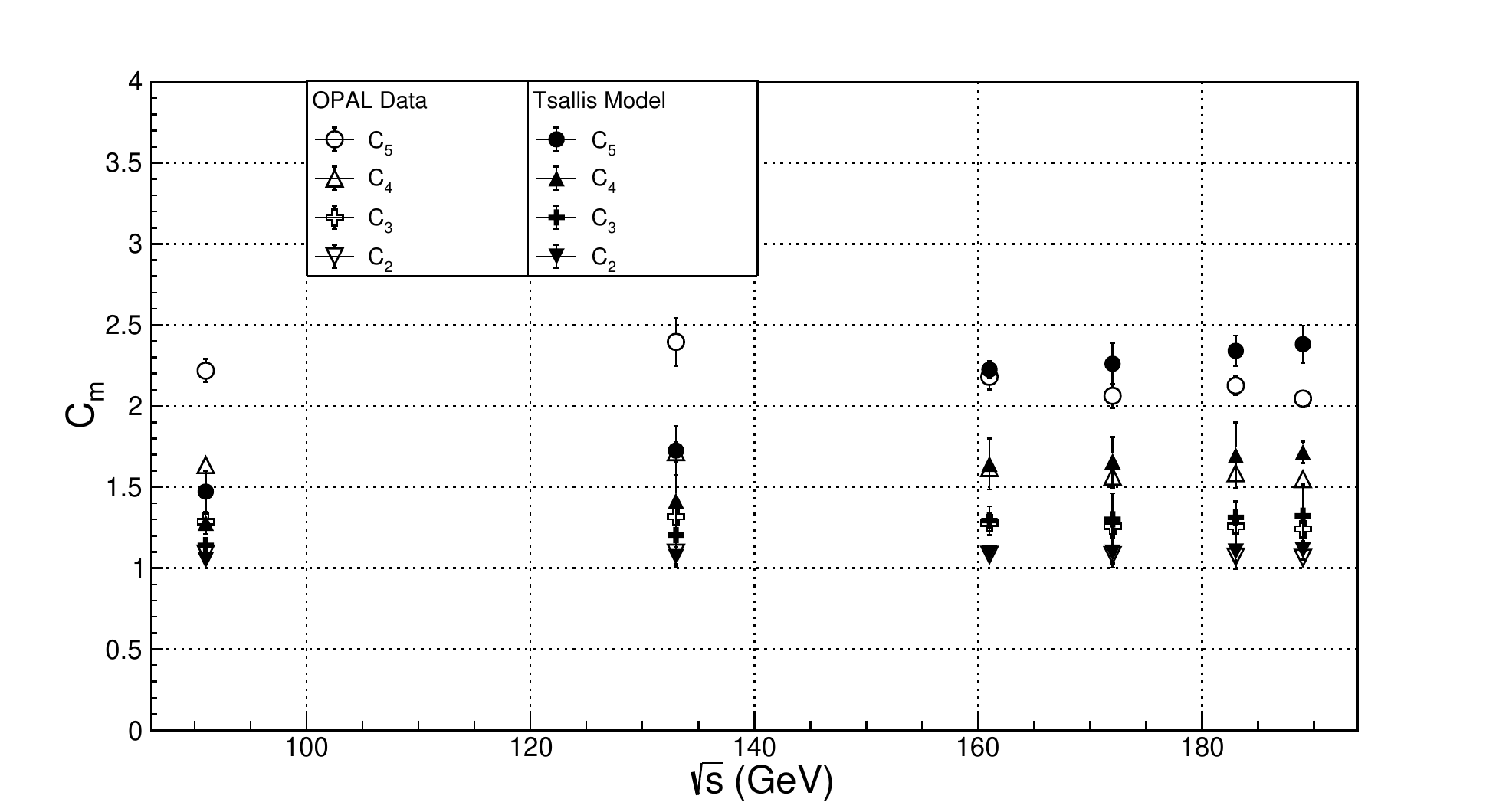}
\includegraphics[width=3.5 in , height= 2.3 in]{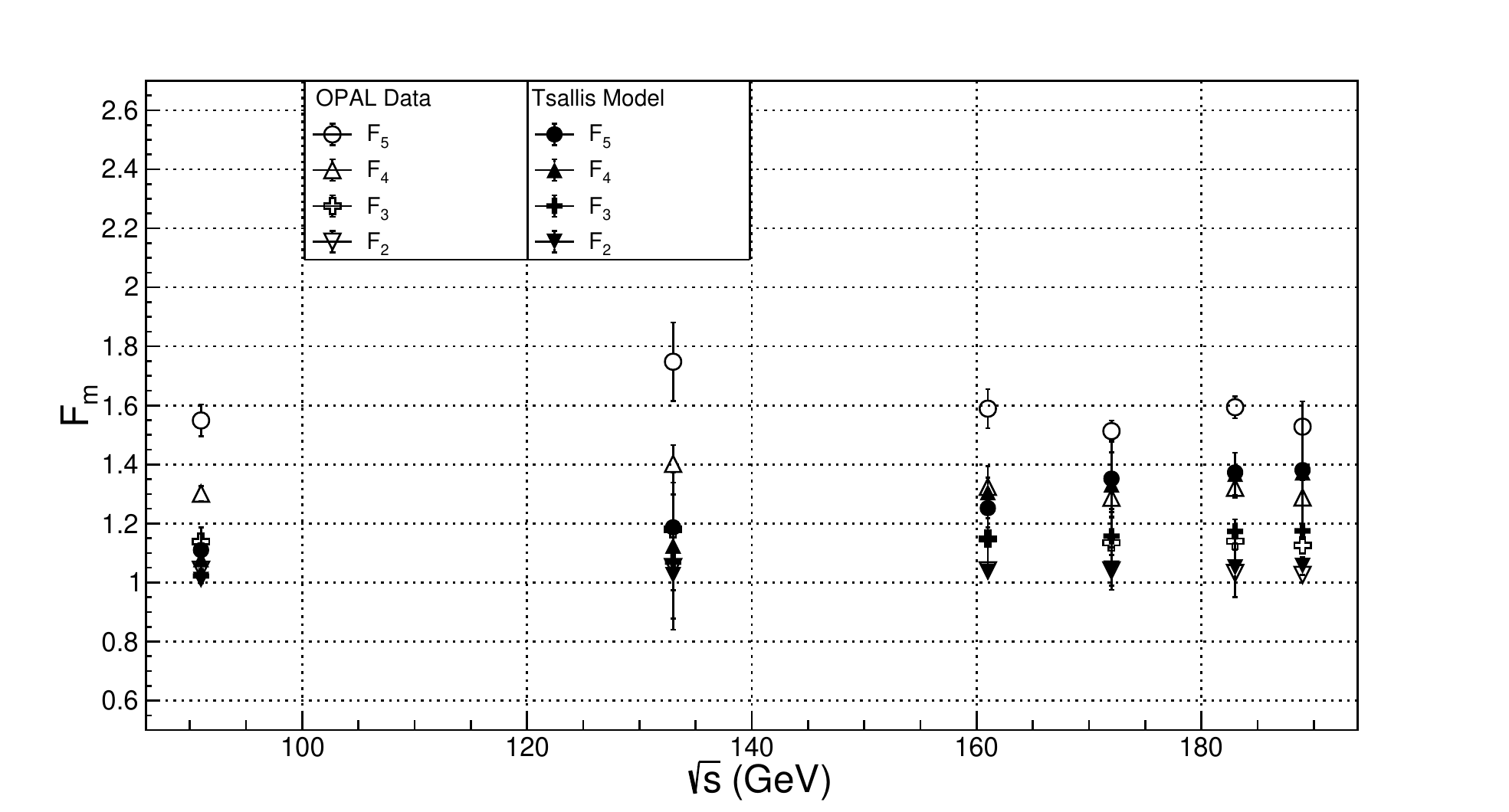}

\caption{Dependence of  $C_{m}$ and $F_{m}$ moments on the centre of mass energy, $\sqrt{s}$  and comparison of the moments obtained using the Tsallis model with the L3 and OPAL experimental values}
\end{figure} 

\begin{figure}[H]
\includegraphics[width=3.5in , height= 2.8 in]{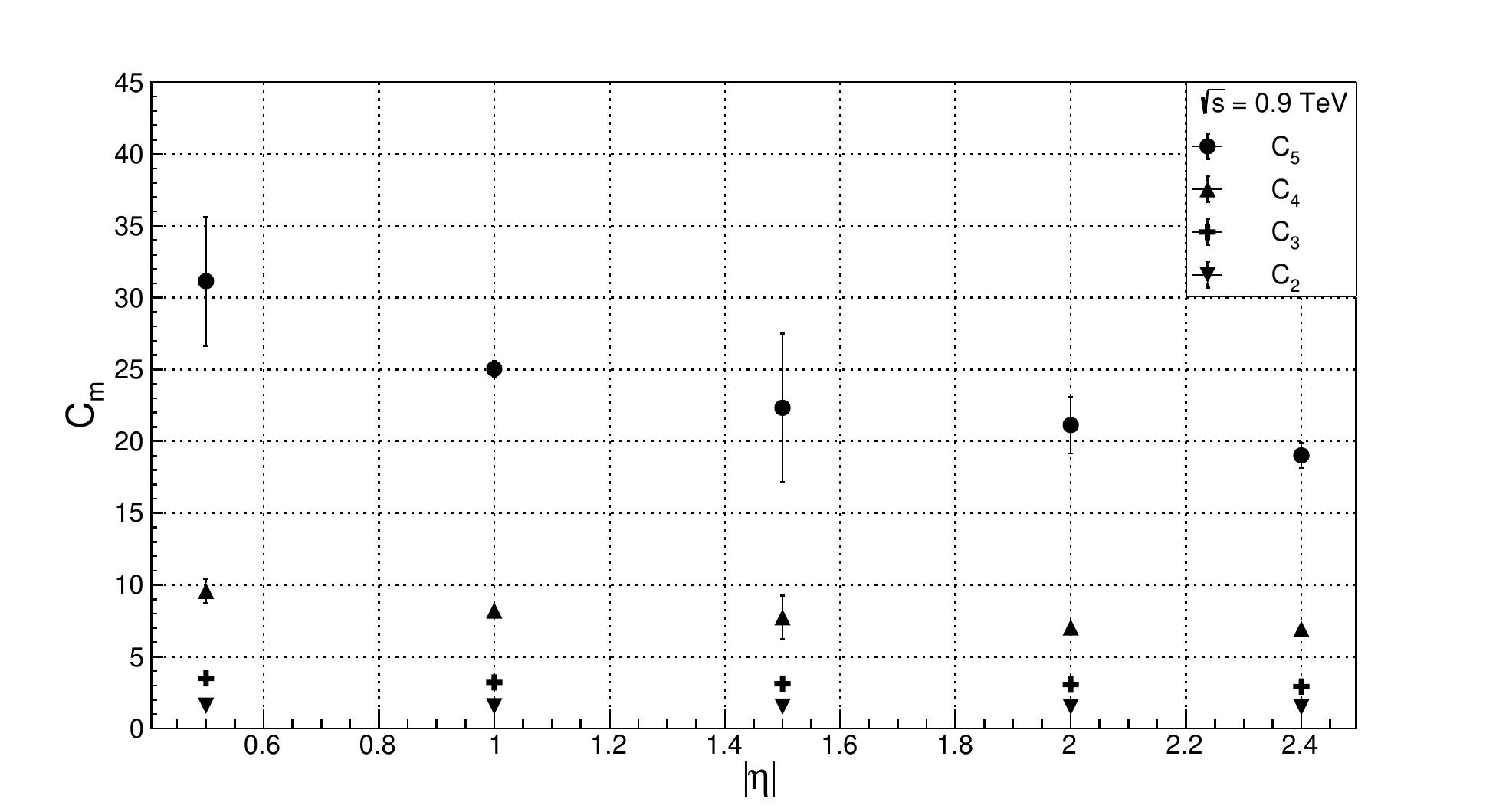}
\includegraphics[width=3.5 in , height= 2.8 in]{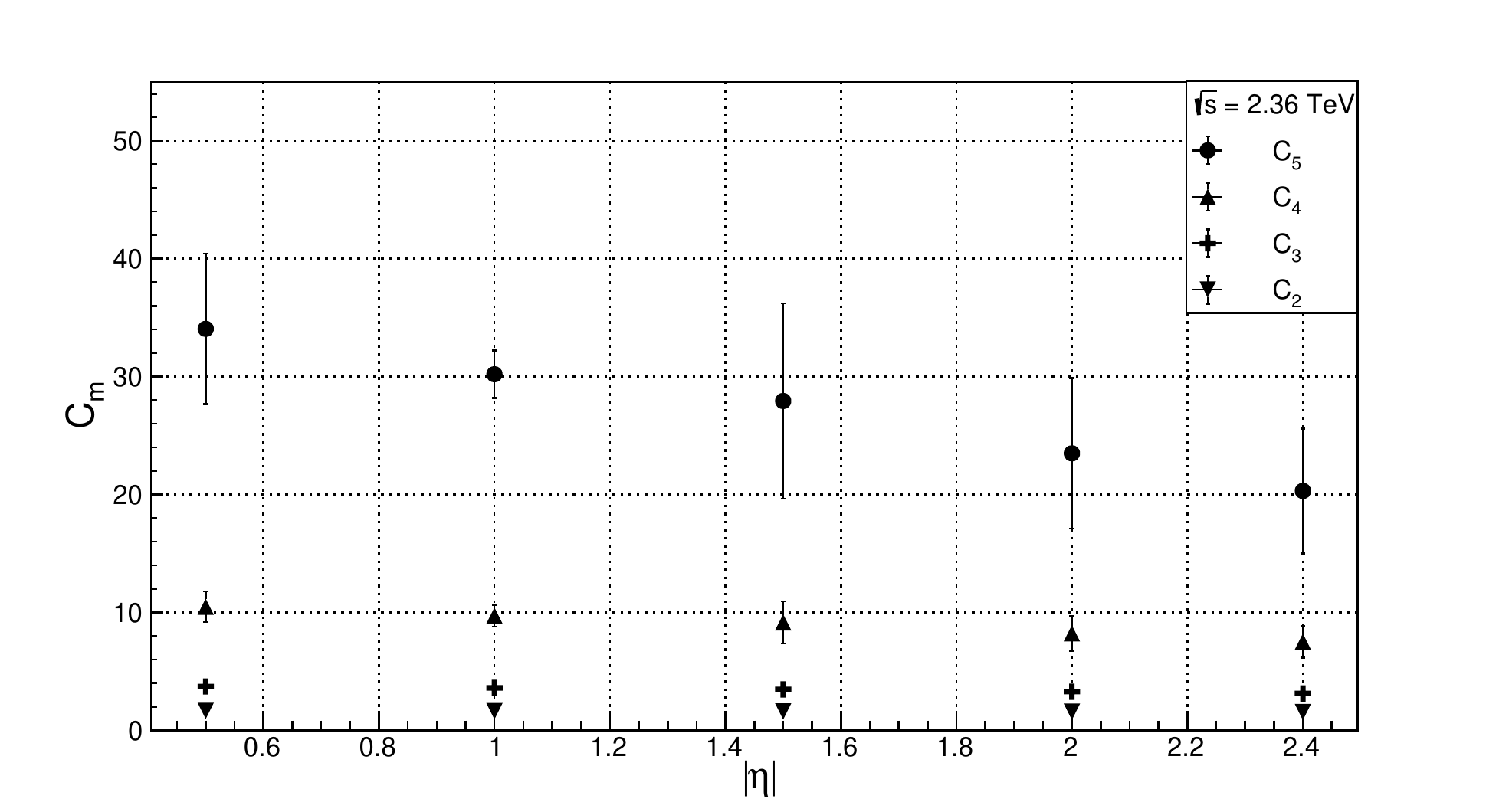}
\includegraphics[width=3.5 in , height= 2.8 in]{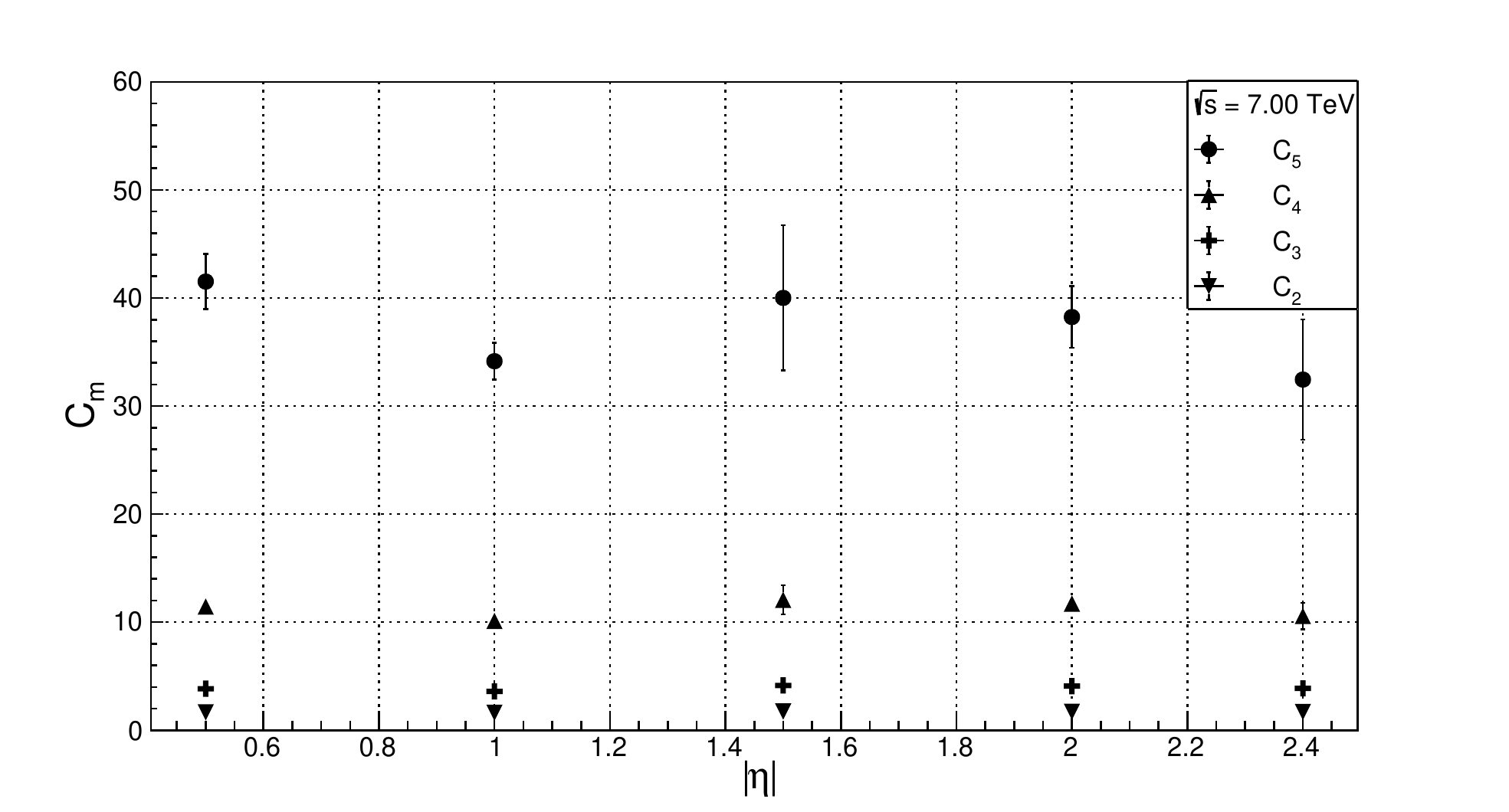}

\caption{$C_{m}$ moments obtained from the Tsallis model and its dependence on  pseudo-rapidity intervals $|\eta|$ at $\sqrt{s}$ = 0.9, 2.36 and 7~TeV for the $pp$ data}
\end{figure}

\begin{figure}[H]

\includegraphics[width=3.5 in , height= 2.8 in]{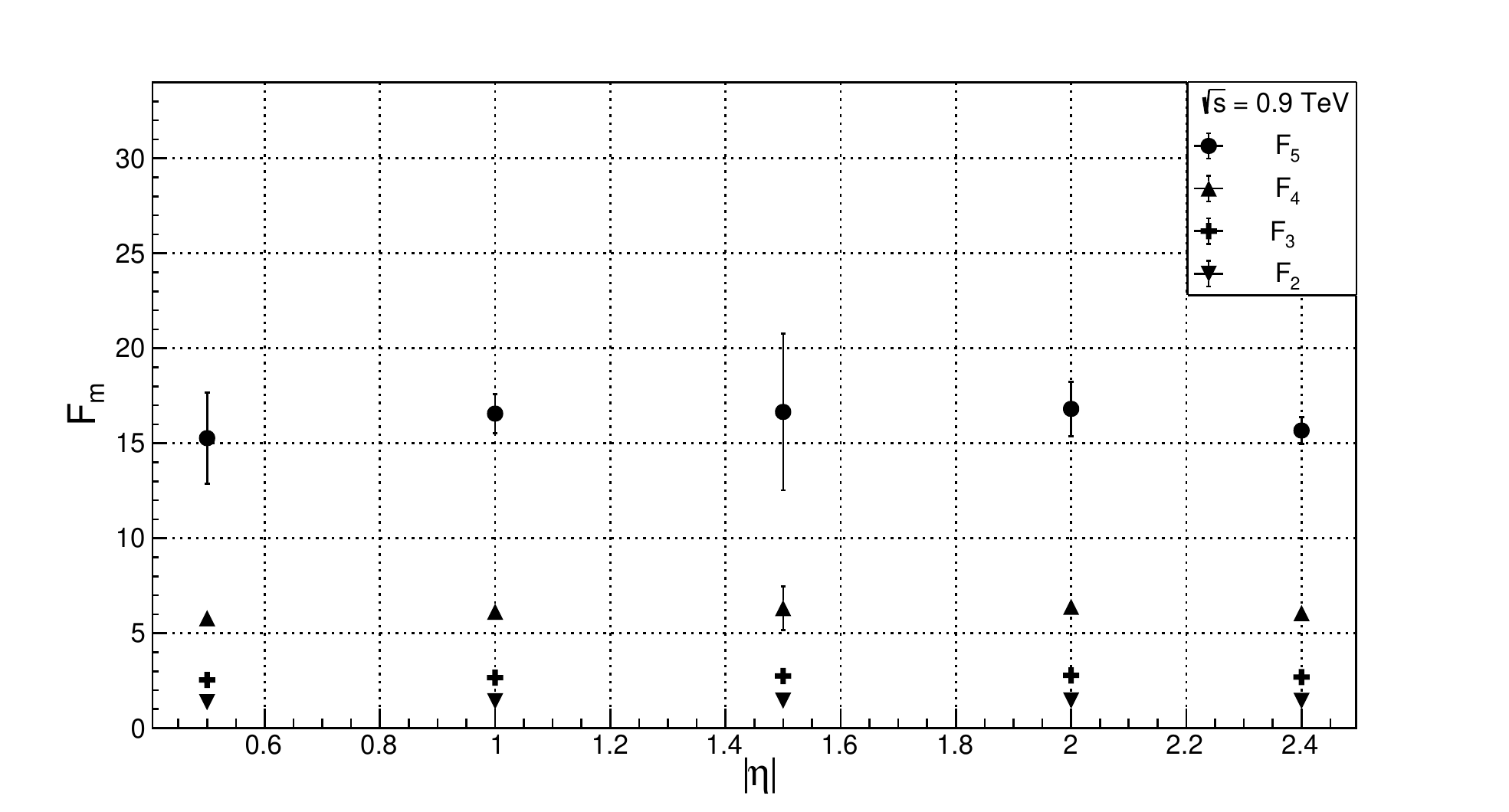}
\includegraphics[width=3.5 in , height= 2.8 in]{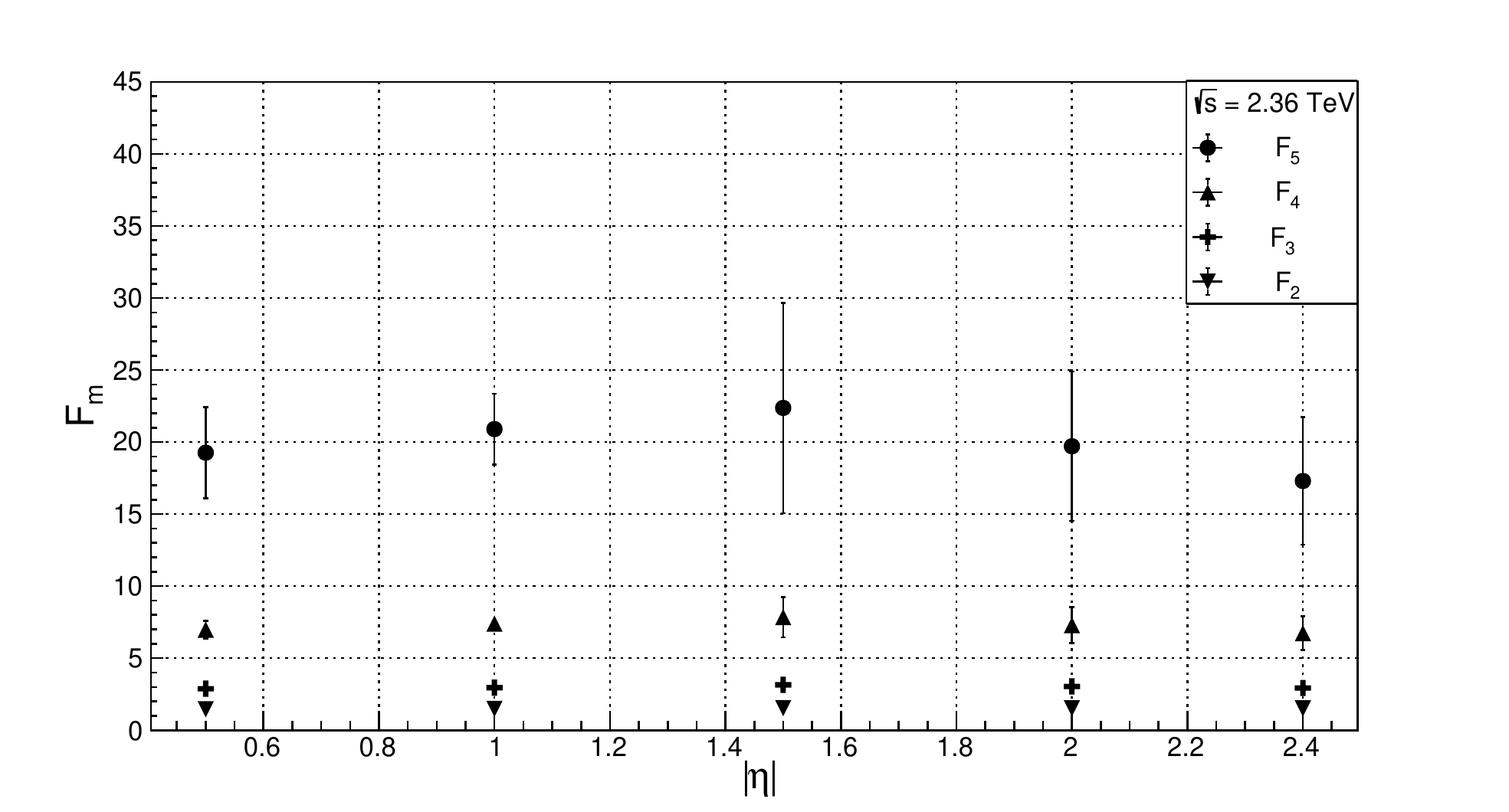}
\includegraphics[width=3.5 in , height= 2.8 in]{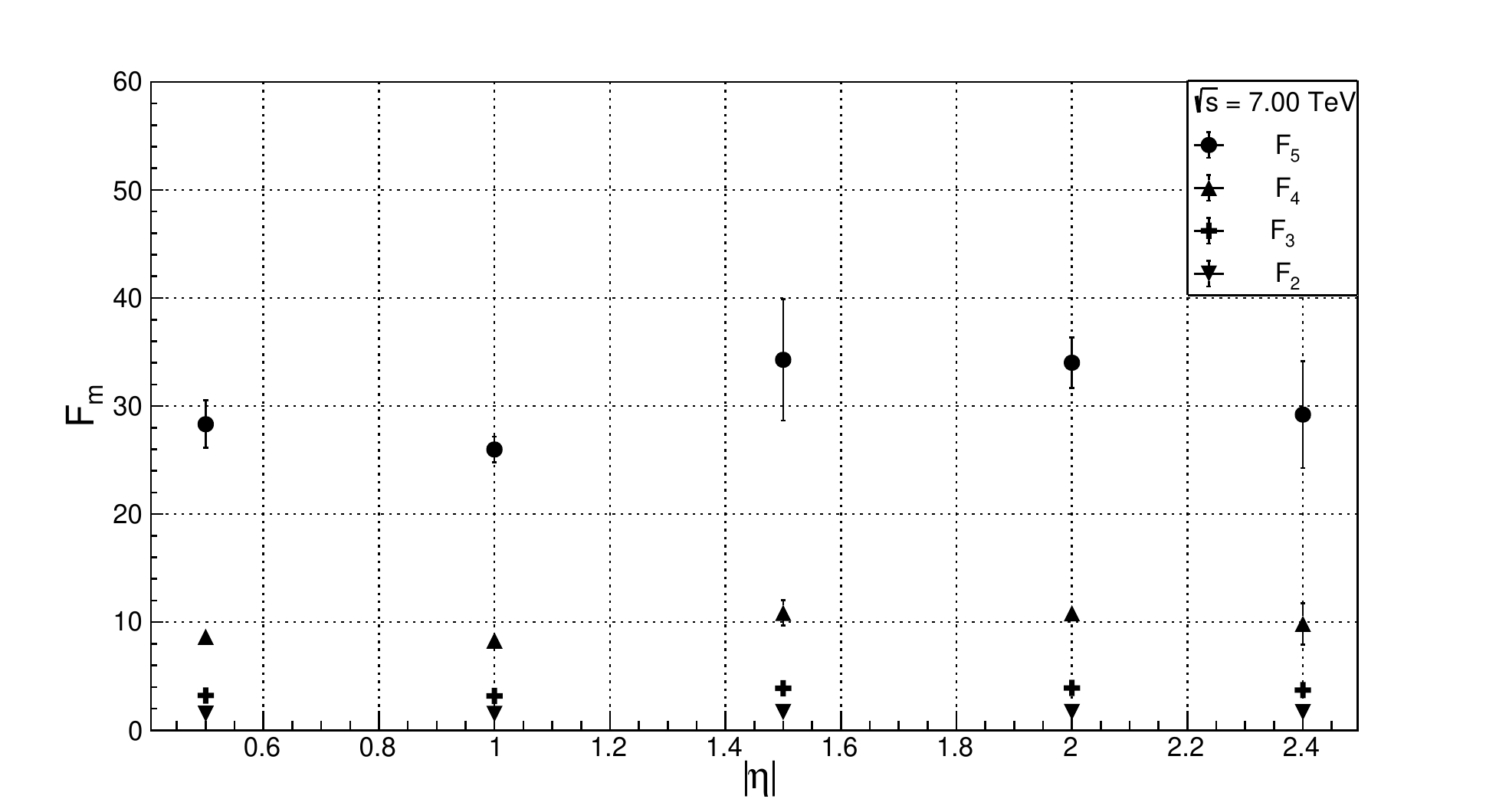}
\caption{$F_{m}$ moments obtained from the Tsallis model and its dependence on  pseudo-rapidity intervals $|\eta|$ at $\sqrt{s}$ = 0.9, 2.36 and 7 TeV for the $pp$ data}
\end{figure}

\begin{figure} [H]
\includegraphics[width=3.5 in , height= 2.9 in]{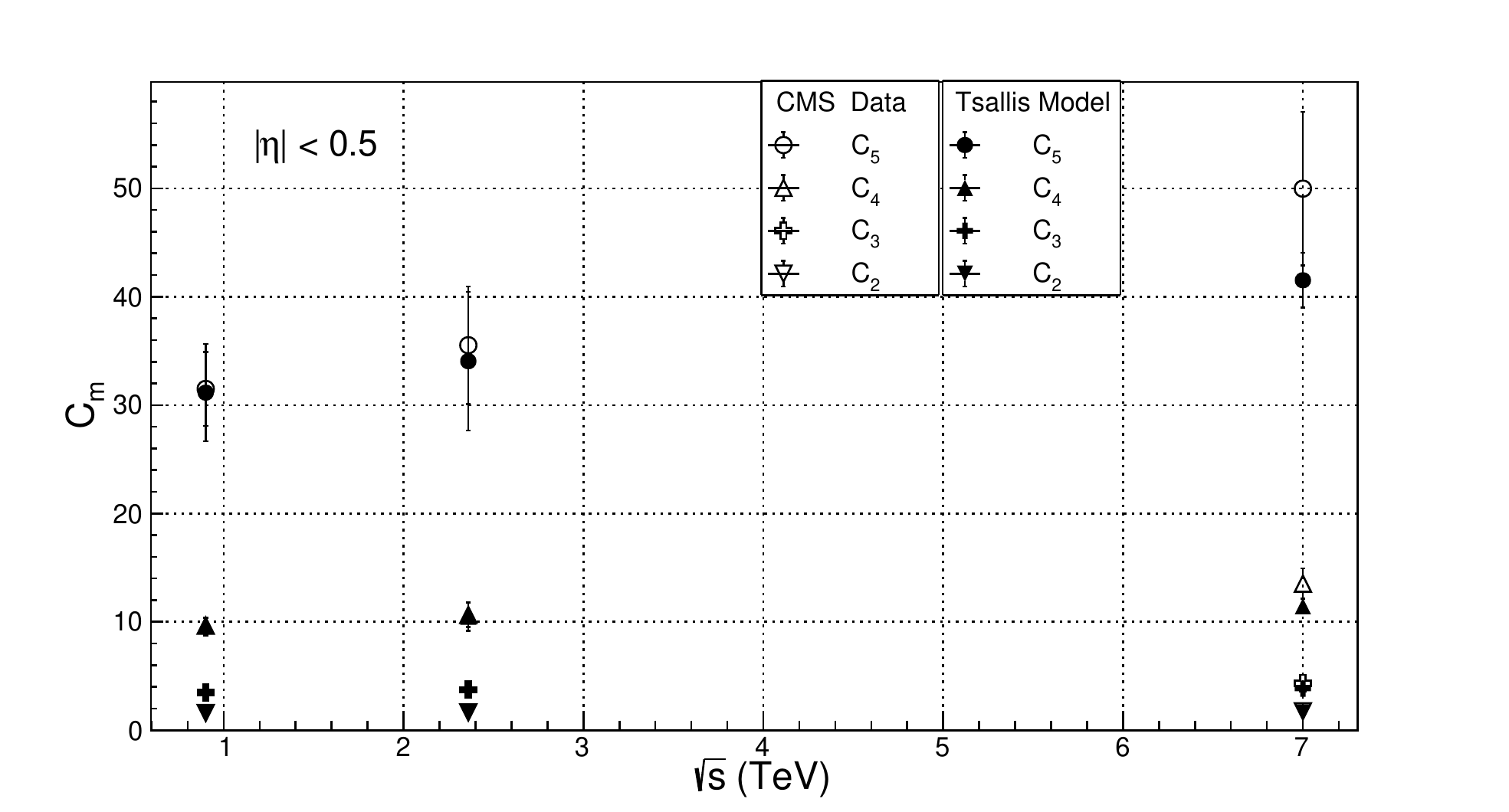}
\includegraphics[width=3.5 in , height= 2.9 in]{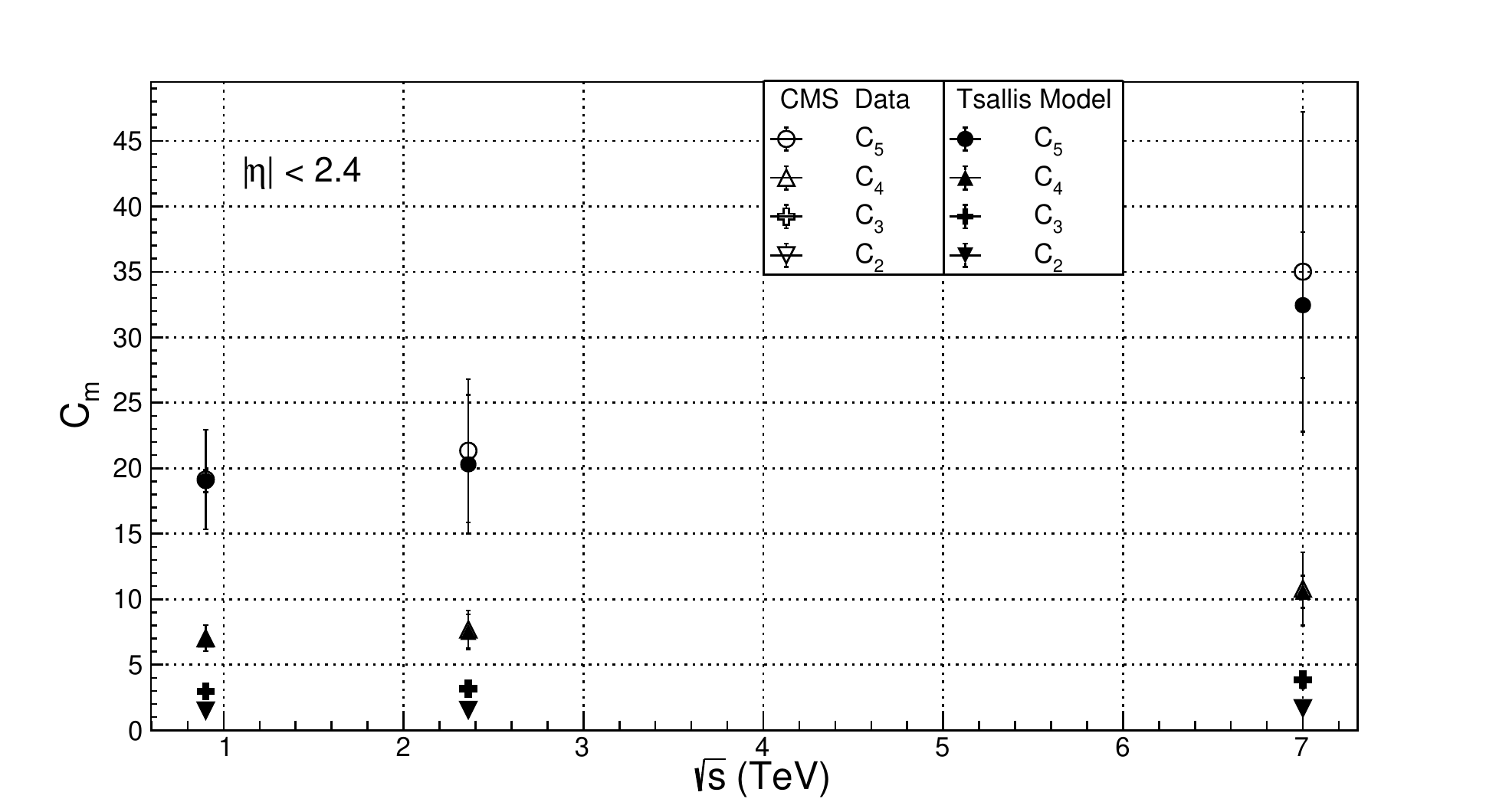}

\caption{The variation of $C_m$ moments  with the centre of mass energy at pseudo-rapidity intervals $|\eta|$ $<$ 0.5 and $|\eta|$ $<$ 2.4 and comparison of the moments calculated from the Tsallis model with the CMS experimental values \cite{cms}}
\end{figure}

\begin{figure} [H]
\includegraphics[width=3.5 in , height= 2.9 in]{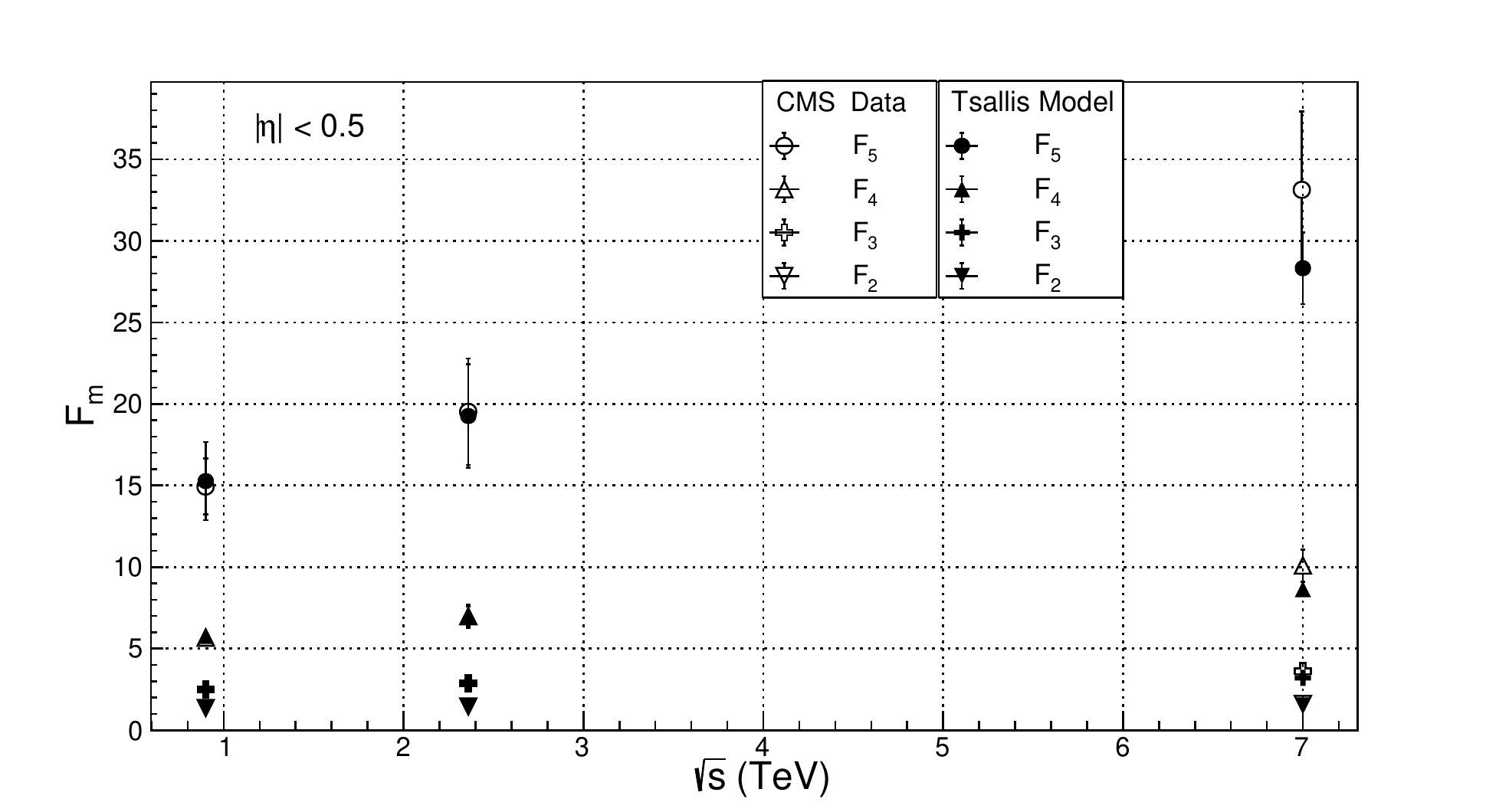}
\includegraphics[width=3.5 in , height= 2.9 in]{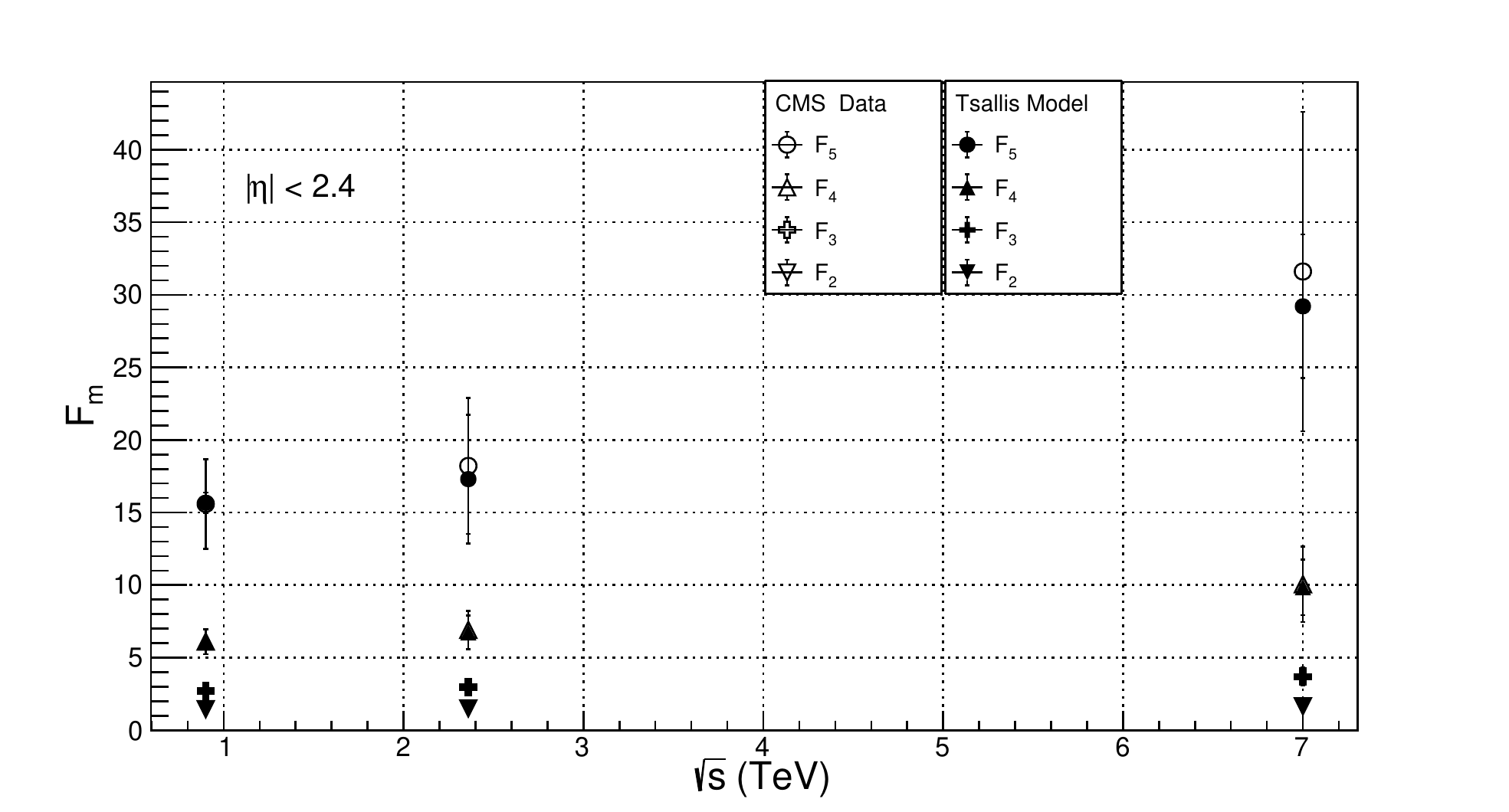}

\caption{The variation of $F_m$ moments  with the centre of mass energy at pseudo-rapidity intervals $|\eta|$ $<$ 0.5 and $|\eta|$ $<$ 2.4 and comparison of the moments calculated from the Tsallis model with the CMS experimental values \cite{cms}}
\end{figure}

%%%%%%%%%%%%%%%%NEW TABLE %%%%%%%%%%%%%%%%%%%

 \begin{table*}
%\begin{center}

%\resizebox{\textwidth}{!}{\begin{tabular}{c|c|cc|cc|cc|cc}
\begin{tabular}{|c|c|cc|}
\hline

  %&\multicolumn{2}{c}{} \\ 
 Experiment & Energy &\multicolumn{2}{|c|}{Average Charged Multiplicity $<N>$ } \\\hline 
 % & &\multicolumn{2}{|c}{} \\ \hline 

   & &&\\

    & & Experimental & \: \: Tsallis  Model   \\
   & & &\\\hline
  %\multicolumn{5}{|c|}{}  \\ 
  		 %\multicolumn{5}{|c|}{$e^+ e^-$ Interactions}  \\       
			% \multicolumn{5}{|c|}{}    \\\hline

   \multirow{6}{*}{OPAL}  &91   & 21.40 $\pm$ 0.43  & 21.07$\pm$ 0.21	 \\\cline{2-4}

	& 133& 23.40 $\pm$ 0.65& 23.17 $\pm$ 0.29\\\cline{2-4}
		 &161&24.46 $\pm$ 0.63& 24.01 $\pm$ 0.47\\\cline{2-4}
		 &172 &25.77 $\pm$ 1.05&24.98 $\pm$ 0.53\\\cline{2-4}
		 &183&26.85 $\pm$ 0.58&26.17 $\pm$ 0.39\\\cline{2-4}
		 &189&26.95 $\pm$ 0.53&26.33 $\pm$ 0.66\\\hline
	%	\multicolumn{3}{c}{}  \\ 
  	%		\multicolumn{3}{c}{L3}  \\       
	%		 \multicolumn{3}{c}{}    \\\hline

\multirow{8}{*}{L3} &130.1&  23.28 $\pm$ 0.26&23.21 $\pm$ 0.35\\\cline{2-4}
		 &136.1& 24.13 $\pm$ 0.29&23.53 $\pm$ 0.17\\\cline{2-4}
		 &172.3& 27.00 $\pm$ 0.58&26.93 $\pm$ 0.25\\\cline{2-4}
		&182.8& 26.84 $\pm$ 0.34&26.77 $\pm$ 0.19\\\cline{2-4}
		 &188.6& 26.84 $\pm$ 0.32&26.51 $\pm$ 0.08\\\cline{2-4}
		 &194.4& 27.14 $\pm$ 0.42&26.87 $\pm$ 0.49\\\cline{2-4}
		 &200.2& 27.73 $\pm$ 0.47&27.09 $\pm$ 0.31\\\cline{2-4}
		 &206.2& 28.09 $\pm$ 0.33&27.38 $\pm$ 0.20\\\hline
 
\end{tabular}
\vspace{0.3cm}
\caption{Average multiplicity $<N>$  at $\sqrt{s}$ = 91 GeV to 206 GeV for $e^+ e^-$ interactions. The values obtained from the Tsallis model are compared with the OPAL and the L3 experimental values}
\end{table*}
%%%%%%%%%%%%%%%%%%%%%%%%%%%%%%%%%%%%%%%%%%%%%%%%%%%%%%%%%%%%%%%%

	\begin{table*}
\begin{center}

%\resizebox{\textwidth}{!}{\begin{tabular}{c|c|cc|cc|cc|cc}
\begin{tabular}{|c|c|cc|}
\hline

 %%&  &\multicolumn{2}{c}{} \\ 
$|\eta|$  Interval & Energy (TeV) &\multicolumn{2}{|c|}{Average Charged Multiplicity ($<N>$)} \\\hline 
%& (TeV) &\multicolumn{2}{|c}{$<n>$} \\ \hline 

 & & & \\

  &  & CMS Experiment & Tsallis Model   \\
& & & \\\hline

  \multirow{3}{*}{0.5}    & 0.9    &  4.355  $\pm$ 0.207	   & 4.583  $\pm$ 0.772	 \\ \cline{2-4}

		& 2.36    &	5.262  $\pm$ 0.250   & 5.489 $\pm$ 0.992	 \\ \cline{2-4}
		& 7.00   &	6.808  $\pm$ 0.335   &	7.409 $\pm$ 1.022 \\ \hline
		
  \multirow{3}{*}{2.4}    & 0.9    &	18.320 $\pm$ 1.273   &	18.957 $\pm$ 1.174 \\ \cline{2-4}

		& 2.36    & 23.166 $\pm$ 1.716	   & 23.524 $\pm$ 1.382	 \\ \cline{2-4}
		& 7.00   &	30.516 $\pm$ 3.660   &	31.231 $\pm$ 3.042 \\ \hline
\end{tabular}
\vspace{0.3cm}
\caption{Average multiplicity $<N>$ at two extreme pseudo-rapidity intervals, $|\eta|$ $<$ 0.5 and $|\eta|$ $<$ 2.4 at $\sqrt{s}$ = 0.9, 2.36 and 7 TeV}
\end{center}
\end{table*}

%%%%%%%%%%Table%%%%%%

\begin{table*}[t]
\begin{tabular}{|c|c|c|c|c|c|c|c|c|c|}
\hline

 &\multicolumn{4}{|c|}{} &\multicolumn{4}{|c|}{}   \\
 Energy  & \multicolumn{4}{|c|}{Experimental Reduced Moments }  &\multicolumn{4}{|c|}{Tsallis Reduced Moments}\\ 
(GeV)  & \multicolumn{4}{|c|}{}  &\multicolumn{4}{|c|}{}\\ \hline
 
  	  		\multicolumn{9}{|c|}{}  \\ 
  			 \multicolumn{9}{|c|}{OPAL Experiment}  \\       
			  \multicolumn{9}{|c|}{}    \\\hline
		&	          &			 &	  &	  &   &   &  &  \\	  
  	   & $C_2$  & $C_3$ &$C_4$ & $C_5$ &  $C_2$ & $C_3$ & $C_4$ & $C_5$ \\ 
  	 	
  	 	  &	          &			 &	  &	  &   &   &  &  \\\hline 
  	 	  91	&1.089 $\pm$ 0.003	&1.287  $\pm$ 0.012	&1.636  $\pm$ 0.029	&2.218 $\pm$ 0.072& 1.048 $\pm$ 0.011		 &	1.141 $\pm$ 0.032    &  1.280 $\pm$ 0.069  & 1.472 $\pm$ 0.127 \\\hline
  	 
133	&1.095 $\pm$ 0.002	&1.317  $\pm$ 0.021	&1.716  $\pm$ 0.063	&2.396 $\pm$ 0.147 &1.068 $\pm$ 0.059&1.204 $\pm$ 0.181&1.416 $\pm$ 0.341&1.725 $\pm$ 0.153 \\\hline
161	&1.082 $\pm$ 0.002	&1.277  $\pm$ 0.010	&1.618  $\pm$ 0.023	&2.180 $\pm$ 0.077&1.093 $\pm$ 0.021&1.293 $\pm$ 0.088&1.643 $\pm$ 0.157&2.225 $\pm$ 0.052 \\\hline
172	&1.080 $\pm$ 0.052	&1.258  $\pm$ 0.061	&1.565  $\pm$ 0.069	&2.063 $\pm$ 0.074 &1.095 $\pm$ 0.091&1.301 $\pm$ 0.161& 1.659 $\pm$ 0.149&2.261 $\pm$ 0.129\\\hline
183	&1.070 $\pm$ 0.026	&1.257  $\pm$ 0.024	&1.586  $\pm$ 0.044	&2.126 $\pm$ 0.058 &1.102 $\pm$ 0.108&1.313 $\pm$ 0.099&1.696 $\pm$ 0.202&2.341 $\pm$ 0.094 \\\hline
189	&1.063 $\pm$ 0.018	&1.241  $\pm$ 0.019	&1.549  $\pm$ 0.019	&2.046 $\pm$ 0.015 &1.110 $\pm$ 0.057&1.323 $\pm$ 0.193&1.715 $\pm$ 0.066&2.382 $\pm$ 0.115 \\\hline

 	\multicolumn{9}{|c|}{}  \\ 
  			 \multicolumn{9}{|c|}{L3 Experiment}  \\       
			  \multicolumn{9}{|c|}{}    \\\hline

130.1	&1.082 $\pm$ 0.014	&1.258 $\pm$  0.057	&1.563 $\pm$  0.042	&2.058 $\pm$  0.096 &1.065 $\pm$ 0.012		 &	1.195 $\pm$ 0.041     &  1.401 $\pm$ 0.096  & 1.705 $\pm$ 0.193  \\\hline
136.1	&1.095 $\pm$ 0.002	&1.301 $\pm$  0.007	&1.656 $\pm$  0.019	&2.237 $\pm$ 0.045&1.069 $\pm$ 0.015&1.208 $\pm$ 0.046& 1.426 $\pm$ 0.112&1.748 $\pm$ 0.226 \\\hline
172.3	&1.094 $\pm$ 0.004	&1.299 $\pm$  0.012	&1.656 $\pm$  0.021	&2.245 $\pm$  0.028 &1.082 $\pm$ 0.002&1.253 $\pm$ 0.071&1.534 $\pm$ 0.061&1.993 $\pm$ 0.023 \\\hline
182.8	&1.091 $\pm$ 0.005	&1.287 $\pm$  0.016	&1.626 $\pm$  0.025	&2.180 $\pm$  0.069 &1.084 $\pm$ 0.031 &1.264 $\pm$ 0.011&1.540 $\pm$ 0.018&1.998 $\pm$ 0.046\\\hline
188.6	&1.086 $\pm$ 0.007	&1.273 $\pm$  0.020	&1.591 $\pm$  0.047	&2.106 $\pm$  0.098 &1.087 $\pm$ 0.011&1.269 $\pm$ 0.013&1.566 $\pm$ 0.027&2.081 $\pm$ 0.043 \\\hline
194.4	&1.093 $\pm$ 0.005	&1.294 $\pm$  0.017	&1.644 $\pm$  0.035	&2.216 $\pm$  0.066 &1.090 $\pm$ 0.073&1.274 $\pm$ 0.007&1.578 $\pm$ 0.016&2.098 $\pm$ 0.038 \\\hline
200.2	&1.093 $\pm$ 0.004	&1.294 $\pm$  0.015	&1.643 $\pm$  0.032	&2.215 $\pm$  0.058 &1.092 $\pm$ 0.052&1.284 $\pm$ 0.015&1.584 $\pm$ 0.029&2.139 $\pm$ 0.034  \\\hline
206.2	&1.091 $\pm$ 0.006	&1.290 $\pm$ 0.016	&1.634 $\pm$  0.035	&2.195  $\pm$  0.067&1.093 $\pm$ 0.009&1.291 $\pm$ 0.036 &1.618 $\pm$ 0.097&2.168 $\pm$ 0.092 \\\hline

\end{tabular}
  \caption{$C_{m}$  moments calculated from the Tsallis model for centre of mass energies, $\sqrt{s}$ = 91 to 206 GeV for the $e^+ e^-$ data}
\end{table*}

%%%%%%%%%%%%%%%%%%%%%%%%%%%%%%%%%%%%%%%%%%%%%%%%%%%%%%%%%%%%%%%%%%%%%%%%%%%%%%%%%%%

\begin{table*}[t]
\begin{tabular}{|c|c|c|c|c|c|c|c|c|c|}
\hline

 &\multicolumn{4}{|c|}{} &\multicolumn{4}{|c|}{}   \\
 Energy  & \multicolumn{4}{|c|}{Experimental Factorial Moments }  &\multicolumn{4}{|c|}{Tsallis Factorial Moments}\\ 
(GeV)  & \multicolumn{4}{|c|}{}  &\multicolumn{4}{|c|}{}\\ \hline
 
  	  		\multicolumn{9}{|c|}{}  \\ 
  			 \multicolumn{9}{|c|}{OPAL Experiment}  \\       
			  \multicolumn{9}{|c|}{}    \\\hline
		&	          &			 &	  &	  &   &   &  &  \\	  
  	   & $F_2$  & $F_3$ &$F_4$ & $F_5$ &  $F_2$ & $F_3$ & $F_4$ & $F_5$ \\ 
  	 	
  	 	  &	          &			 &	  &	  &   &   &  &  \\\hline 
  	 	  
91	&1.043 $\pm$ 0.003	&1.139 $\pm$ 0.011	&1.301 $\pm$ 0.026	&1.549 $\pm$ 0.054 &1.009 $\pm$ 0.018		 &	1.022 $\pm$ 0.024    &  1.079 $\pm$ 0.048  & 1.110 $\pm$ 0.077 \\\hline
133	&1.052 $\pm$ 0.002	&1.181 $\pm$ 0.024	&1.402 $\pm$ 0.064	&1.748 $\pm$ 0.133 & 1.025 $\pm$ 0.051& 1.069 $\pm$ 0.229& 1.125 $\pm$ 0.247& 1.187 $\pm$ 0.010 \\\hline
161	&1.041 $\pm$ 0.002	&1.148 $\pm$ 0.010	&1.324 $\pm$ 0.031	&1.589 $\pm$ 0.066 &1.036 $\pm$ 0.019& 1.145 $\pm$ 0.103& 1.306 $\pm$ 0.088& 1.252 $\pm$ 0.065\\\hline
172	&1.041 $\pm$ 0.052	&1.135 $\pm$ 0.043	&1.287 $\pm$ 0.049	&1.513 $\pm$ 0.036 &1.047 $\pm$ 0.072&1.157 $\pm$ 0.092& 1.332 $\pm$ 0.109& 1.352 $\pm$ 0.133\\\hline
183	&1.032 $\pm$ 0.025	&1.140 $\pm$ 0.029	&1.321 $\pm$ 0.033	&1.594 $\pm$ 0.037 &1.051 $\pm$ 0.101&1.173 $\pm$ 0.041& 1.368 $\pm$ 0.072&1.373 $\pm$ 0.014\\\hline
189	&1.026 $\pm$ 0.018	&1.126 $\pm$ 0.017	&1.288 $\pm$ 0.012	&1.528 $\pm$ 0.004 &1.056 $\pm$ 0.031&1.175  $\pm$ 0.011& 1.373 $\pm$ 0.026&1.381 $\pm$ 0.233 \\\hline
 	\multicolumn{9}{|c|}{}  \\ 
  			 \multicolumn{9}{|c|}{L3 Experiment}  \\       
			  \multicolumn{9}{|c|}{}    \\\hline

130.1	&1.039 $\pm$ 0.005	&1.123 $\pm$ 0.018	&1.261 $\pm$ 0.041	&1.465 $\pm$ 0.084 &1.034 $\pm$ 0.009		 &	1.101 $\pm$ 0.072    &  1.199 $\pm$ 0.061  & 1.329 $\pm$ 0.117 \\\hline
136.1	&1.054 $\pm$ 0.002	&1.167 $\pm$ 0.008	&1.352 $\pm$ 0.018	&1.624 $\pm$ 0.040 &1.041 $\pm$ 0.009& 1.086 $\pm$ 0.134& 1.162 $\pm$ 0.082&1.388 $\pm$ 0.133\\\hline
172.3	&1.057 $\pm$ 0.005	&1.181 $\pm$ 0.014	&1.384 $\pm$ 0.025	&1.691 $\pm$ 0.035 &1.045 $\pm$ 0.003& 1.142 $\pm$ 0.006&1.278 $\pm$ 0.014&1.502 $\pm$ 0.023\\\hline
182.8	&1.053 $\pm$ 0.003	&1.167 $\pm$ 0.017	&1.354 $\pm$ 0.026	&1.633 $\pm$ 0.064 &1.048 $\pm$ 0.010& 1.145 $\pm$ 0.009 & 1.291 $\pm$ 0.010&1.515 $\pm$ 0.035\\\hline
188.6	&1.049 $\pm$ 0.005	&1.154 $\pm$ 0.019	&1.322 $\pm$ 0.048	&1.571 $\pm$ 0.089&1.050 $\pm$ 0.021& 1.146 $\pm$ 0.008& 1.301 $\pm$ 0.003&1.522 $\pm$ 0.061\\\hline
194.4	&1.056 $\pm$ 0.006	&1.176 $\pm$ 0.018	&1.374 $\pm$ 0.036	&1.670 $\pm$ 0.061&1.051 $\pm$ 0.005&1.152 $\pm$ 0.046& 1.307 $\pm$ 0.017&1.547 $\pm$ 0.009\\\hline
200.2	&1.057 $\pm$ 0.005	&1.178 $\pm$ 0.016	&1.378 $\pm$ 0.032	&1.676 $\pm$ 0.056 &1.053 $\pm$ 0.013&1.158 $\pm$ 0.011&1.319 $\pm$  0.021&1.552 $\pm$ 0.030\\\hline
206.2	&1.056 $\pm$ 0.006	&1.175 $\pm$ 0.018	&1.372 $\pm$ 0.035	&1.665 $\pm$ 0.063 &1.056 $\pm$ 0.012&1.164 $\pm$ 0.052&1.334 $\pm$ 0.033&1.572 $\pm$ 0.039\\\hline

\end{tabular}
  \caption{$F_{m}$  moments calculated from the Tsallis model for centre of mass energies, $\sqrt{s}$ = 91 to 206 GeV for the $e^+ e^-$ data}
\end{table*}

%%%%%%%%%%%%%%%%%%%%%%%%%%%%%%%%%%%%%%%%

\begin{table*}[t]
\begin{tabular}{|c|c|c|c|c|c|c|c|c|c|}
\hline

 &\multicolumn{4}{|c|}{} &\multicolumn{4}{|c|}{}   \\
 Pseudo-rapidity  & \multicolumn{4}{|c|}{Tsallis Reduced Moments }  &\multicolumn{4}{|c|}{Tsallis Factorial Moments}\\ 
Interval  & \multicolumn{4}{|c|}{}  &\multicolumn{4}{|c|}{}\\ \hline
 
  	  		\multicolumn{9}{|c|}{}  \\ 
  			 \multicolumn{9}{|c|}{$\sqrt{s}$ = 0.9 TeV }  \\       
			  \multicolumn{9}{|c|}{}    \\\hline
		&	          &			 &	  &	  &   &   &  &  \\	  
  	  $|\eta|$ & $C_2$  & $C_3$ &$C_4$ & $C_5$ &  $F_2$ & $F_3$ & $F_4$ & $F_5$ \\ 
  	 	
  	 	  &	          &			 &	  &	  &   &   &  &  \\\hline 
0.5	& 1.59 $\pm$ 0.02	& 3.49 $\pm$ 0.14	& 9.59 $\pm$ 0.83 & 31.15 $\pm$ 4.49	& 1.37 $\pm$ 0.01& 2.54 $\pm$ 0.01& 5.79 $\pm$ 0.21 & 15.27 $\pm$ 2.40\\\hline
1.0	  &	1.55 $\pm$ 0.02		 &	3.21 $\pm$ 0.10     &  8.22 $\pm$ 0.35  & 25.04 $\pm$ 0.55  & 1.42 $\pm$ 0.01& 2.66 $\pm$ 0.06& 6.13 $\pm$ 0.25& 16.56 $\pm$ 1.02\\\hline
 
 1.5	  &	1.53 $\pm$ 0.06		 & 3.12 $\pm$ 0.35	     &  7.75 $\pm$ 1.52 &  22.33 $\pm$ 5.17 & 1.45 $\pm$ 0.04 & 2.75 $\pm$ 0.24 & 6.32 $\pm$ 1.15 & 16.65 $\pm$ 4.12\\\hline
 
 2.0	  &	1.51 $\pm$ 0.03		 & 3.06 $\pm$ 0.10	     & 7.05 $\pm$ 0.44   &   21.13 $\pm$ 1.97 & 1.46 $\pm$ 0.02& 2.78 $\pm$ 0.08& 6.39 $\pm$ 0.34 & 16.81 $\pm$ 1.43\\\hline

2.4	  &	1.49 $\pm$ 0.01		 &	2.92 $\pm$ 0.05     & 6.93 $\pm$ 0.22   & 19.02 $\pm$ 0.85   & 1.44 $\pm$ 0.01 & 2.69 $\pm$ 0.05& 6.06 $\pm$ 0.19 & 15.67 $\pm$ 0.70\\\hline
 	\multicolumn{9}{|c|}{}  \\ 
  			 \multicolumn{9}{|c|}{$\sqrt{s}$ = 2.36 TeV}  \\       
			  \multicolumn{9}{|c|}{}    \\\hline

0.5	  &	1.64 $\pm$ 0.03		 & 3.71 $\pm$ 0.23	     &  10.48 $\pm$ 1.29 & 34.05 $\pm$ 6.39   & 1.45 $\pm$ 0.01& 2.88 $\pm$ 0.09& 6.98 $\pm$ 0.62 & 19.26 $\pm$ 3.16\\\hline
 
 1.0	  &		1.62 $\pm$ 0.04	 & 3.59 $\pm$ 0.29	     & 9.73 $\pm$ 0.93    & 30.20 $\pm$ 2.01   & 1.48 $\pm$ 0.02 & 2.95 $\pm$ 0.07 & 7.40 $\pm$ 0.32 & 20.90 $\pm$ 2.46\\\hline
 
 1.5	  &		1.60 $\pm$ 0.06	 & 3.46 $\pm$ 0.36	     & 9.15 $\pm$ 1.78   & 27.93 $\pm$ 8.29   &  1.53 $\pm$ 0.04& 3.15 $\pm$ 0.29& 7.85 $\pm$ 1.40& 22.37 $\pm$ 7.30\\\hline
 
 2.0	  & 1.57 $\pm$ 0.05			 &	 3.27 $\pm$ 0.32     & 8.22 $\pm$ 1.49   & 23.49 $\pm$ 6.38   & 1.52 $\pm$ 0.05 & 3.04 $\pm$ 0.27 & 7.29 $\pm$ 1.25& 19.71 $\pm$ 5.19\\\hline

 2.4	  &	1.55 $\pm$ 0.01		 & 3.12 $\pm$0.31	     & 7.51 $\pm$ 1.34   & 20.30 $\pm$ 5.30   & 1.51 $\pm$ 0.06 & 2.93 $\pm$ 0.27 & 6.74 $\pm$ 1.16 & 17.30 $\pm$ 4.42 \\\hline

  	\multicolumn{9}{|c|}{}  \\ 
  			 \multicolumn{9}{|c|}{$\sqrt{s}$ = 7 TeV}  \\       
			  \multicolumn{9}{|c|}{}    \\\hline

  	0.5	  &	1.65 $\pm$ 0.02		 & 3.84 $\pm$ 0.10	     & 11.44 $\pm$ 0.48   & 41.52 $\pm$ 2.54   & 1.52 $\pm$ 0.02 & 3.21 $\pm$ 0.08 & 8.65 $\pm$ 0.37 & 28.32 $\pm$ 2.20 \\\hline

 1.0	 &	1.62 $\pm$ 0.02		 & 3.59 $\pm$ 0.07	     & 10.11 $\pm$ 0.34   & 34.15 $\pm$ 1.70   & 1.52 $\pm$ 0.01 & 3.17 $\pm$ 0.05 & 8.30 $\pm$ 0.24 & 25.98 $\pm$ 1.18\\\hline
 
  1.5	  &	1.73 $\pm$ 0.04		 &	4.14 $\pm$ 0.26     & 12.07 $\pm$ 1.35   & 40.01 $\pm$ 6.72   & 1.68 $\pm$ 0.03  & 3.88 $\pm$ 0.22 & 10.86 $\pm$ 1.17& 34.28 $\pm$ 5.63\\\hline
 
  2.0	  &	1.71 $\pm$ 0.01		 &	4.08 $\pm$ 0.14     & 11.72 $\pm$ 0.51  &   38.24 $\pm$ 2.84 & 1.69 $\pm$ 0.02 & 3.91 $\pm$ 0.14 & 10.82 $\pm$ 0.52 & 34.01 $\pm$ 2.35\\\hline

 2.4	  &	1.69 $\pm$ 0.04		 & 3.87 $\pm$ 0.25	     & 10.57 $\pm$ 1.22   & 32.45 $\pm$ 5.57   & 1.67 $\pm$ 0.05 & 3.71 $\pm$ 0.23 & 9.85 $\pm$ 1.92 & 29.21 $\pm$ 4.95\\\hline

\end{tabular}
  \caption{$C_{m}$ and $F_{m}$ moments calculated from the Tsallis model for different pseudo-rapidity intervals at centre of mass energy, $\sqrt{s}$ = 0.9, 2.36 and 7 TeV for the $pp$ data}
\end{table*}

%%%%%%%%%%%%%%%%%%%%%%%%%%%%%%%%%%%%%%%%%%%%%%%%%%%%%%%%%%%%%%%%%%

%%%%%%%%%%%%%%%%%%%%%%%%%%%%%%%%%%%%%%%%%%%%%%%%%%%%%%%%%%%%%%%%%%%%

%

\end{document}